\documentclass[%
reprint,
superscriptaddress,
 amsmath,amssymb,
 aps,
prd,
]{revtex4-2}

\usepackage{graphicx}% Include figure files
\usepackage{dcolumn}% Align table columns on decimal point
\usepackage{bm}% bold math
\usepackage{amssymb}
\usepackage{pifont}
\usepackage{xcolor}

\begin{document}

%\preprint{APS/123-QED}

\title{Enhancing noise characterization with robust time delay interferometry combination}
% Force line breaks with \\

\author{Gang Wang}
\email[Gang Wang: ]{gwang@shao.ac.cn, gwanggw@gmail.com}
%\thanks{}
\affiliation{Shanghai Astronomical Observatory, Chinese Academy of Sciences, Shanghai 200030, China}

\date{\today}% It is always \today, today,

\begin{abstract}

Time delay interferometry (TDI) is essential for suppressing laser frequency noise and achieving the targeted sensitivity for space-borne gravitational wave (GW) missions. In Paper I, we examined the performance of the fiducial second-generation TDI Michelson configuration versus an alternative, the hybrid Relay, in noise suppression and data analysis. The results showed that both TDI schemes have comparable performances in mitigating laser and clock noises. However, when analyzing chirp signal from the coalescence of massive binary black holes, the Michelson configuration becomes inferior due to its vulnerable T channel and numerous null frequencies. In contrast, the hybrid Relay is more robust in dynamic unequal-arm scenarios. In this work, we further investigate the noise characterization capabilities of these two TDI configurations. Our investigations demonstrate that hybrid Relay achieves more robust noise parameter inference than the Michelson configuration. Moreover, the performance can be enhanced by replacing the T channel of hybrid Relay with a second-generation TDI null stream $C^{12}_3$. The combined three data streams, including two science observables from the hybrid Relay and $C^{12}_3$, could reduce the instabilities of noise spectra in the targeting frequency band and form an optimal dataset for characterizing noises.

\end{abstract}

\keywords{Gravitational Wave, Time-Delay Interferometry, LISA}

\maketitle

\section{Introduction}

LISA is scheduled to be launched in the 2030s and is designed to observe gravitational waves (GWs) in the mHz frequency band \cite{2017arXiv170200786A,Colpi:2024xhw}. Three spacecraft (S/C) will form a $2.5 \times 10^6$ km triangular constellation with six interferometric links. Time delay interferometry (TDI) was developed to suppress the dominant laser frequency noise and achieve the targeted sensitivity for space-borne interferometers \cite{1997SPIE.3116..105N,1999ApJ...527..814A,2000PhRvD..62d2002E}. The principle of TDI is to combine the interferometric laser links with proper delays to form equivalent equal-arm interferometry. The first-generation TDI was proposed to cancel the laser noise in a static unequal-arm interferometer \cite{1999ApJ...527..814A}. Due to orbital dynamics, second-generation TDI is required to mitigate the effects caused by relative motions between spacecraft (S/C) \cite{Tinto:2003vj,Shaddock:2003dj}.

For each TDI configuration, three observables can be obtained by choosing an initial S/C and reordering path sequences based on the same geometry. Three optimal TDI observables, (A, E, T), could be derived by applying an orthogonal transformation \cite{Prince:2002hp,Vallisneri:2007xa}.
Since each TDI observable utilizes different interferometric arms with different time delays, they will yield varying capacities for laser noise suppression and sensitivities \cite{Wang:2020pkk}.
The second-generation Michelson configuration, as a fiducial TDI scheme, is widely used to perform data analysis assuming an equilateral triangular constellation. 
However, the performance of Michelson can diverge from the equal-arm case under a realistic unequal-arm scenarios.
Laser noise cancellations and sensitivities in the unequal-arm case were investigated for the first-generation TDI ordinary channels \cite{Larson:2002xr,Cornish:2003tz}. As for the optimal TDI observables, \citet{Adams:2010vc} found sensitivity divergence of Michelson-T between equal and unequal arm cases in the low frequency band. \citet{Hartwig:2023pft} investigated this improvement of T channel, which is fully correlated with E channel. Furthermore, Wang \textit{et al.} \cite{Wang:2020fwa,Wang:2020a} evaluated the sensitivities of the first-generation TDI in dynamic scenarios and found that both noise spectrum and response function of Michelson-T channel are sensitive to the variations in unequal-arm lengths.
\citet{Katz:2022yqe} assessed the bias in inferring parameters of semi-monochromatic sources introduced by treating the unequal arms as equal.

In recent works, we employed the second-generation TDI Michelson configuration to analyze the chirp signals from massive black holes coalescences and found the instability of T channel undermines the analysis \cite{Wang:2024ssp}. Additionally, the symmetry of the geometry introduces numerous null frequencies in the Michelson observables. To mitigate these two disadvantages of Michelson configuration, we proposed the hybrid Relay TDI scheme to perform data analysis in Paper I \cite{Wang:2024alm}. The alternative TDI observables are less sensitive to changes in arm lengths, and their science channels have only one quarter of the null frequencies compared to Michelson, which reduce the suppression of the GW signal.

In this work, we further investigate the capability of hybrid Relay in noise characterization and compare it with the fiducial Michelson. The results show that the T channel is crucial for breaking the degeneracy between optical metrology system (OMS) noises in the science channels and for precisely determining noise parameters. The performance of Michelson configuration is still limited by its unstable T channel and numerous null frequencies, whereas the hybrid Relay shows potential for superior capability. However, the performance of hybrid Relay is also affected by its T observable, which is subject to increased null frequencies. To enhance its capability, the null stream $C^{12}_3$ identified in \cite{Hartwig:2021mzw} is selected to replace the hybrid Relay-T. The combined three data streams, A and E of the hybrid Relay and $C^{12}_3$, can reduce the instability of noise spectra in the targeting frequency band and improve efficiency of noise characterization.

This paper is organized as follows: In Section \ref{sec:tdi}, we introduce the second-generation TDI configurations used in this investigation. 
Section \ref{sec:correlations} evaluates the cross-correlations between the TDI observables for both noise spectra and GW responses, and compares the stabilities of TDI spectra in a dynamic orbit.
The noise characterizations are performed using simulated data in Section \ref{sec:simulation_characterization}. The different capabilities in inferring noise parameters reveal the disadvantages of the Michelson configuration and the advantages of the hybrid Relay.
A brief conclusion and discussion are given in Section \ref{sec:conclusions}.
(We set $G=c=1$ in this work except where specified otherwise in the equations.)

\section{Time delay interferometry} \label{sec:tdi}

\begin{figure}[htb]
\includegraphics[width=0.20\textwidth]{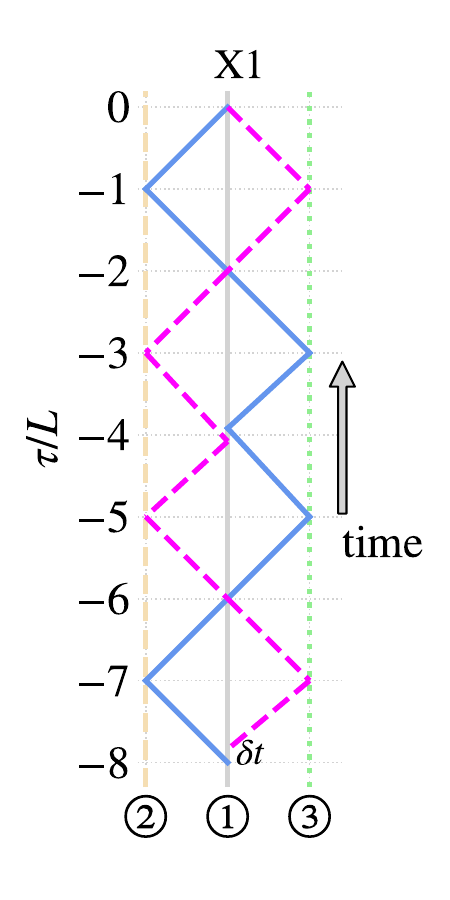}
\includegraphics[width=0.27\textwidth]{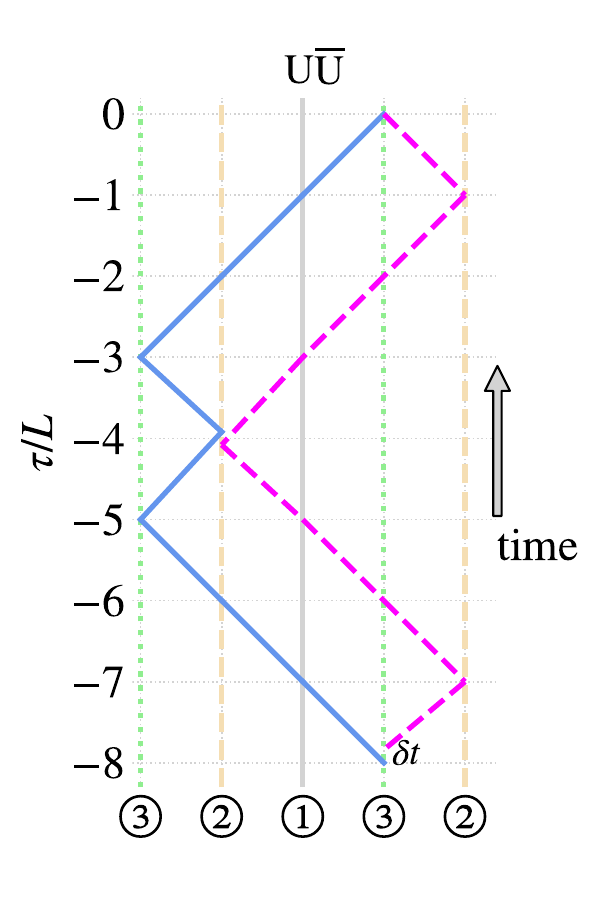}
\caption{\label{fig:2nd_tdi_diagram} The geometric diagrams for X1 and $\mathrm{U\overline{U}}$ \cite{Wang:2011,Wang:2020pkk} (diagrams reused from \cite{Wang:2024alm}). The vertical lines show the trajectories of spacecraft in the time direction, and the \textcircled{$i$} indicates S/C$i$ ($i = 1, 2, 3$). Additional trajectories of S/C2 and S/C3 are plotted in $\mathrm{U\overline{U}}$ to avoid the crossings at noninteger delay time points. The solid blue lines show path of one laser beam, and the dashed magenta lines show the path of another laser beam. Two beams form the equivalent interference in TDI.}
\end{figure}

Most classical second-generation TDI can be constructed by synthesizing the first-generation TDI or employing other methods \cite{Vallisneri:2005ji,Hartwig:2021mzw,Tinto:2022zmf,Wang:2020pkk}. In the case of the fiducial second-generation Michelson TDI, each observable utilizes four laser interferometric links from two arms. By selecting different initial S/C and sequence, three observables (X1, Y1, Z1) can be defined,
\begin{eqnarray} \label{eq:X1_measurement}
{\rm X1}  &= \mathcal{B}_{121313121}  -   \mathcal{B}_{131212131}, \\
{\rm Y1}  &= \mathcal{B}_{232121232}  -   \mathcal{B}_{212323212}, \\
{\rm Z1}  &= \mathcal{B}_{313232313}  -   \mathcal{B}_{323131323},
\end{eqnarray}
where $\mathcal{B}$ is a beam expression along a path, and its subscripts from right to left indicate the S/C indexes in temporal sequence. For instance, the first beam of $\mathrm{X1}$, $\mathcal{B}_{121313121} $, indicates the blue solid beam in the left geometry diagram of Fig. \ref{fig:2nd_tdi_diagram}. A measurements along the beams S/C1$\rightarrow$2$\rightarrow$1$\rightarrow$3$\rightarrow$1 will be expressed as
\begin{equation}
\mathcal{B}_{13121} = \eta_{13} +  \mathcal{D}_{13} \eta_{31} +  \mathcal{D}_{13} \mathcal{D}_{31} \eta_{12} + \mathcal{D}_{13} \mathcal{D}_{31} \mathcal{D}_{12} \eta_{21},
\end{equation}
where $\mathcal{D}_{ij}$ is a delay operator defined as $\mathcal{D}_{ij} y = y(t - L_{ij})$, $L_{ij}$ is the ranging from S/C$j$ to S/C$i$, and $\eta_{ij}$ represents the measurement combination in a laser interferometric link from S/C$j$ to S/C$i$ as specified in \cite{Otto:2012dk,Otto:2015,Wang:2024ssp}. 

Three observables of hybrid Relay, ($\mathrm{U\overline{U}, \ V\overline{V}, \ W\overline{W}}$), can be formulated as
\begin{eqnarray}
\mathrm{U\overline{U}} = & \mathcal{B}_{312323213} - \mathcal{B}_{323121323}, \\
\mathrm{V\overline{V}} = & \mathcal{B}_{123131321} - \mathcal{B}_{131232131}, \\
\mathrm{W\overline{W}} =& \mathcal{B}_{231212132} - \mathcal{B}_{212313212}.
\end{eqnarray}
The geometry of $\mathrm{U\overline{U}} $ is illustrated in the right diagram of Fig. \ref{fig:2nd_tdi_diagram} \cite{Wang:2011,Wang:2020pkk}. Three channels of second-generation TDI Sagnac, ($\alpha1$, $\beta1$, $\gamma1$), could be represented as
\begin{eqnarray}
{\rm \alpha 1}  =&  \mathcal{B}_{1231321}  -  \mathcal{B}_{1321231}, \\
{\rm \beta 1}  =&  \mathcal{B}_{2312132}  -  \mathcal{B}_{2132312}, \\
{\rm \gamma 1} =& \mathcal{B}_{3123213}  -  \mathcal{B}_{3213123}.
\end{eqnarray}
These three TDI configurations -- Michelson, hybrid Relay, and Sagnac -- should all meet the requirements for suppressing laser noise and be eligible to perform the GW analysis \cite{Wang:2011,Wang:2020pkk}. For brevity, we refer to TDI observables as second-generation in this study and without explicitly emphasizing the \textit{second-generation}.

\begin{figure}[thb]
\includegraphics[width=0.48\textwidth]{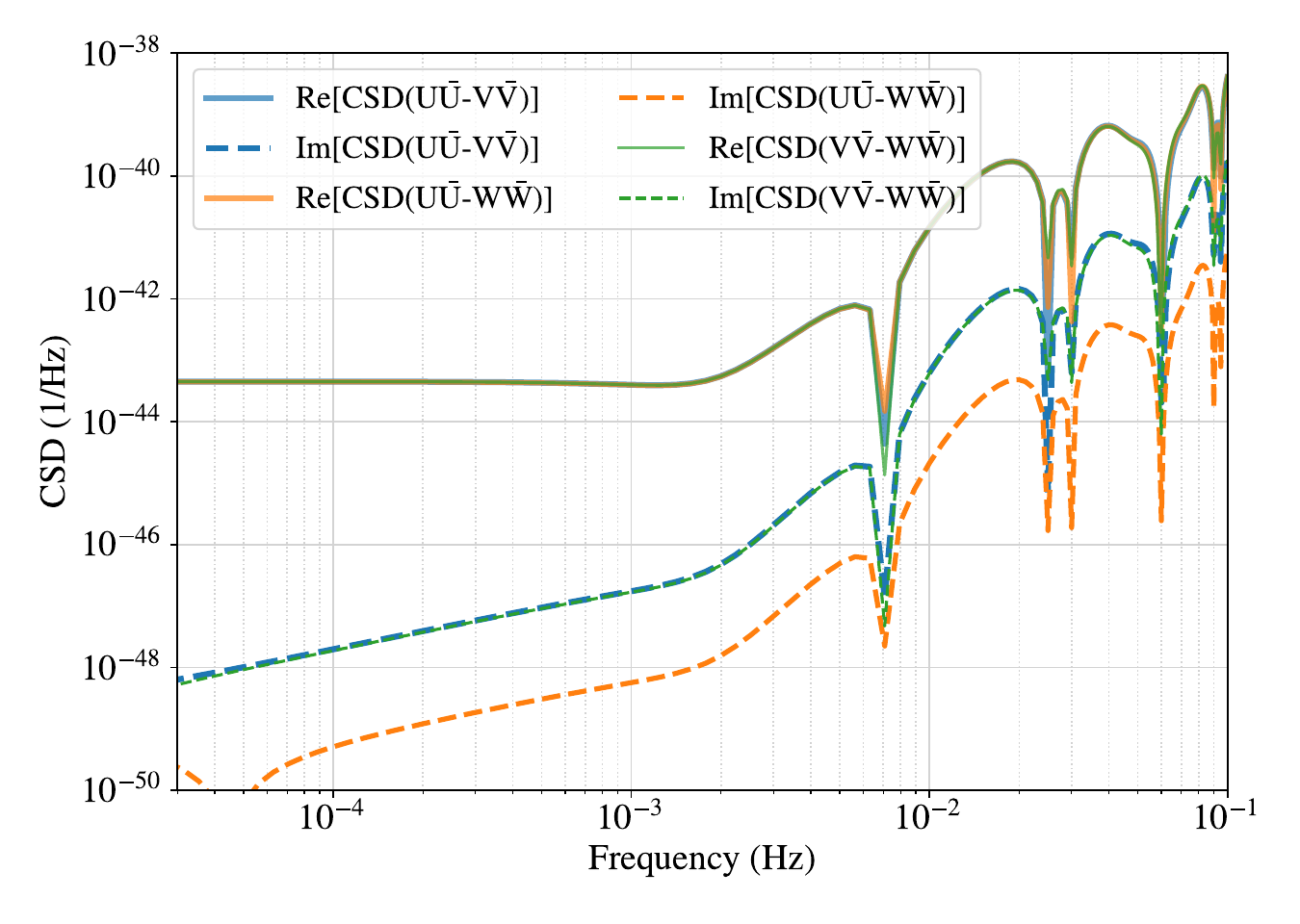} 
\caption{\label{fig:noise_csd_Relay} The real and imaginary components of CSDs for hybrid Relay observables. The real parts of CSDs dominate over imaginary components, enabling orthogonal transform from (U$\mathrm{\bar{U}}$, V$\mathrm{\overline{V}}$, W$\mathrm{\overline{W}}$) to (A$_\mathrm{U\overline{U}}$, E$_\mathrm{U\overline{U}}$, T$_\mathrm{U\overline{U}}$). (The real components from three pairs are overlapped.) }
\end{figure}

The covariance matrix of noises from three ordinary channels $(a, b, c)$ of a TDI configuration, such as the three ordinary channels of Michelson (X1, Y1, Z1), can be expressed as
\begin{equation}
\begin{bmatrix}
S_\mathrm{a} & S_\mathrm{ab} & S_\mathrm{ac} \\
S_\mathrm{ba} & S_\mathrm{b} & S_\mathrm{bc} \\
S_\mathrm{ca} & S_\mathrm{cb} & S_\mathrm{c}
\end{bmatrix}
\simeq 
\begin{bmatrix}
S_\mathrm{a} & S_\mathrm{ab} & S_\mathrm{ab} \\
S_\mathrm{ab} & S_\mathrm{a} & S_\mathrm{ab} \\
S_\mathrm{ab} & S_\mathrm{ab} & S_\mathrm{a}
\end{bmatrix}, 
\end{equation}
where $S_\mathrm{a}$ represents the power spectral density (PSD) of data stream $a$, and $S_\mathrm{ab}$ denotes cross spectral density (CSD) of data stream $a$ and $b$.  The $(a, b, c)$ could also represent the triple ordinary channels ($\alpha1$, $\beta1$, $\gamma1$) for Sagnac or ($\mathrm{U\overline{U}, \ V\overline{V}, \ W\overline{W}}$) for hybrid Relay.
The (quasi-)orthogonal or optimal observables (A, E, T) can then be derived from the eigenvectors of the covariance matrix under the condition that all CSDs of data streams are equal in principle \cite{Prince:2002hp,Vallisneri:2007xa},
\begin{equation} \label{eq:abc2AET}
\begin{bmatrix}
\mathrm{A}_a  \\ \mathrm{E}_a  \\ \mathrm{T}_a
\end{bmatrix}
 =
\begin{bmatrix}
-\frac{1}{\sqrt{2}} & 0 & \frac{1}{\sqrt{2}} \\
\frac{1}{\sqrt{6}} & -\frac{2}{\sqrt{6}} & \frac{1}{\sqrt{6}} \\
\frac{1}{\sqrt{3}} & \frac{1}{\sqrt{3}} & \frac{1}{\sqrt{3}}
\end{bmatrix}
\begin{bmatrix}
a \\ b  \\ c
\end{bmatrix}.
\end{equation}
In realistic scenario, the CSDs of $S_{ab}$ and $S_{ba}$ are conjugate. However, the orthogonal transform still holds (closely) when the real part of CSD dominates over the imaginary part. The CSDs across the hybrid Relay observables are shown in Fig. \ref{fig:noise_csd_Relay}, where the plot demonstrates that the real components are orders of magnitude higher than their imaginary components. Therefore, the transform from ordinary observables ($\mathrm{U\overline{U}, \ V\overline{V}, \ W\overline{W}}$) to the optimal observables (A$_\mathrm{U\overline{U}}$, E$_\mathrm{U\overline{U}}$, T$_\mathrm{U\overline{U}}$) is applicable. To distinguish the optimal observables from different TDI configurations, a subscript of the first ordinary observable is added to A/E/T. For example, (A$_\mathrm{X1}$, E$_\mathrm{X1}$, T$_\mathrm{X1}$) are orthogonal observables from Michelson (X1, Y1, Z1). A checklist for TDI observables is provided in Table \ref{tab:tdi_list}.

\begin{table}[tbh]
\caption{\label{tab:tdi_list} Checklist of observables for second-generation TDI configurations: Michelson, hybrid Relay, and Sagnac.}
\begin{ruledtabular}
\begin{tabular}{c|cc}
TDI configuration & ordinary  & optimal  \\
\hline
Michelson &  (X1, Y1, Z1) & (A$_\mathrm{X1}$, E$_\mathrm{X1}$, T$_\mathrm{X1}$) \\
hybrid Relay & ($\mathrm{U\overline{U}, \ V\overline{V}, \ W\overline{W}}$) & (A$_\mathrm{U\overline{U}}$, E$_\mathrm{U\overline{U}}$, T$_\mathrm{U\overline{U}}$) \\
Sagnac & ($\alpha1$, $\beta1$, $\gamma1$) &  (A$_\mathrm{\alpha1}$, E$_\mathrm{\alpha1}$, T$_\mathrm{\alpha1}$)
\end{tabular}
\end{ruledtabular}
\end{table}

For the optimal data streams, A and E can effectively respond to GW signals, with their antenna patterns equivalent to two orthogonal interferometer rotated by $\pi/4$ \cite{Wang:2020fwa}. The T channel will be a noise-dominated data stream, especially at frequencies lower than 50 mHz. As detailed in the following section, the T channel plays a crucial role in noise characterization as null stream.

\begin{figure}[htb]
\includegraphics[width=0.4\textwidth]{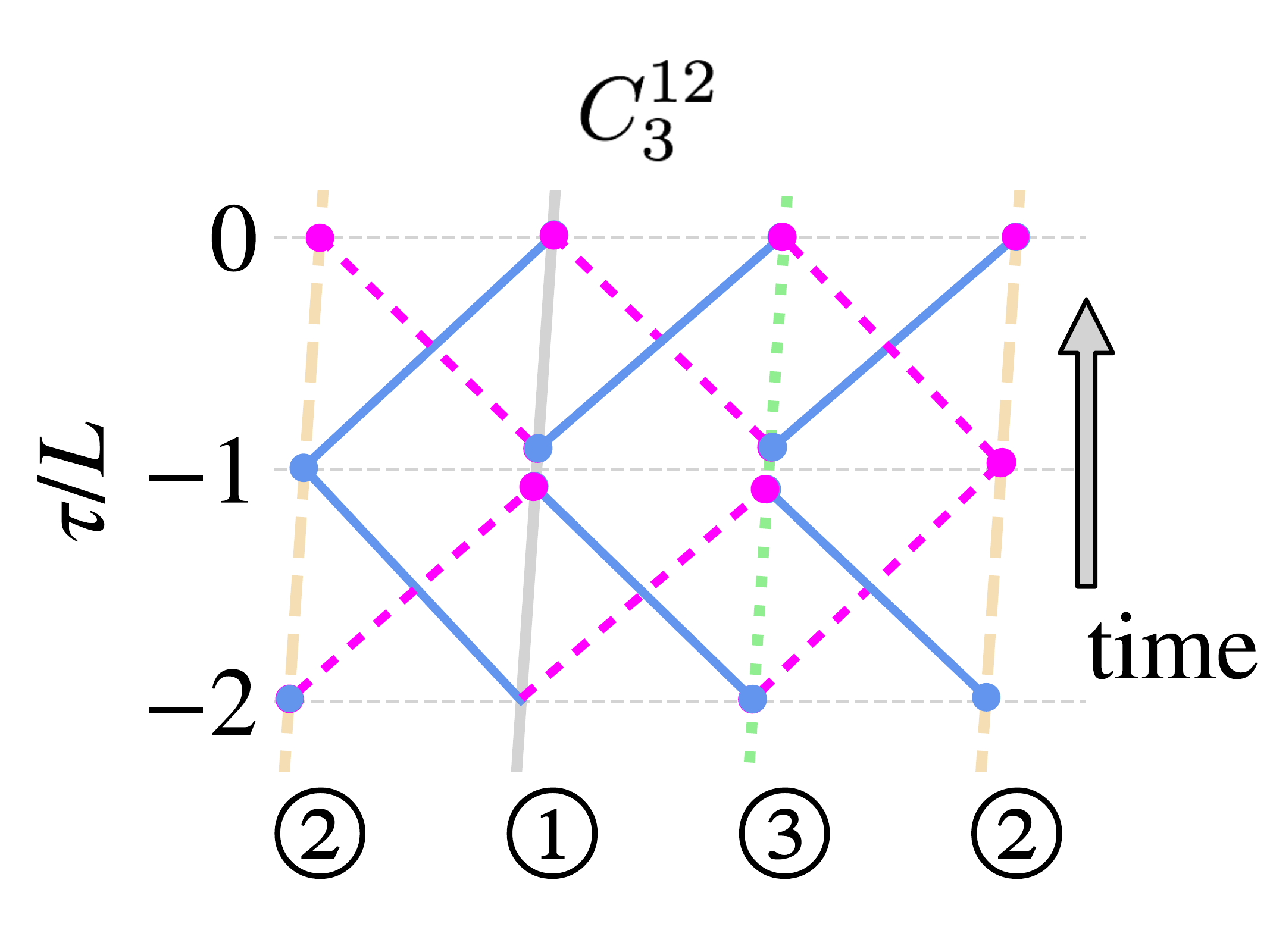}
\caption{\label{fig:null_tdi_diagram} The geometric diagram for $C^{12}_3$. Additional trajectory of S/C2 (\textcircled{2}) is plotted. The trajectories of the spacecrafts are deliberately tilted to illustrate rotation effect of constellation, which results in $L_{ij} (t) \neq L_{ji}(t)$ and yields the Sagnac effect. As described in \cite{Hartwig:2021mzw,Hartwig:2021dlc}, the closed-loop path of $C^{12}_3$ starts and ends at S/C1 at $\tau=-2L$, and the first link is from S/C1 to S/C2 along the solid blue line. The path switches at at each joint depending on the subsequence color.}
\end{figure}

Besides the T channels by combining three ordinary channels, the specific null streams have been developed. Typical cases include the first-generation TDI fully symmetric Sagnac $\zeta$ \cite{1999ApJ...527..814A} and the the second-generation TDI $\zeta_1$ \cite{Tinto:2003vj}. In the unequal arms case, replacing T channel in the first-generation TDI Michelson with $\zeta$ could enhance noise characterization and stochastic GW reconstruction \cite{Hartwig:2023pft}.
\citet{Hartwig:2021mzw} developed more second-generation TDI null streams and and examined their noise spectra. Among these null streams, $C^{12}_3$ has null frequencies at $i/L \ (i= 1,2,3...)$, and $C^{16}_{28}$ has no null frequency in target frequency band. Therefore, these two null observables are selected as potential substitutes for the T channels. The geometries of the $C^{12}_3$ are shown in Fig. \ref{fig:null_tdi_diagram}.

\section{Correlations between TDI and spectra stability} \label{sec:correlations}

\begin{figure*}[thb]
\includegraphics[width=0.48\textwidth]{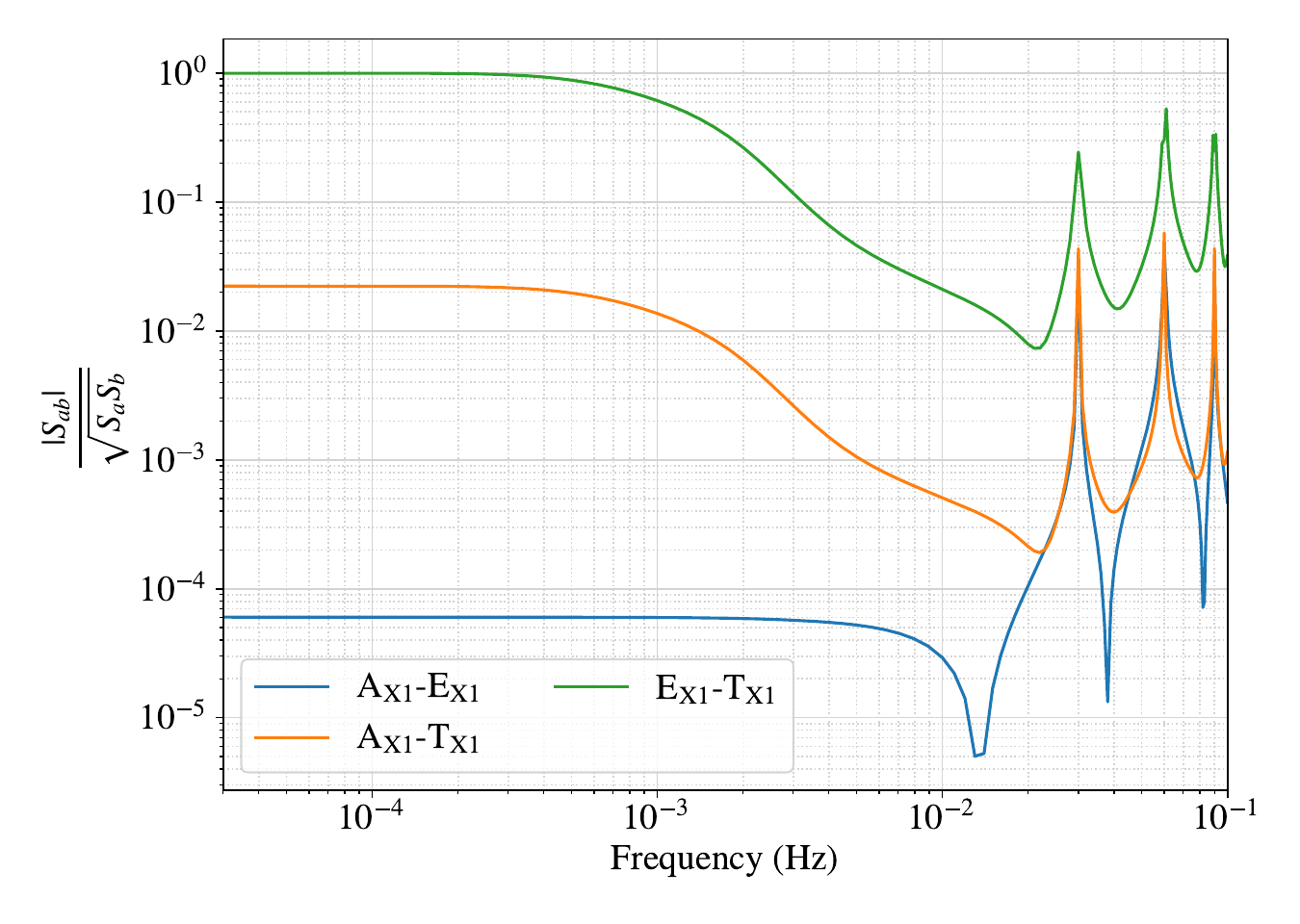}
\includegraphics[width=0.48\textwidth]{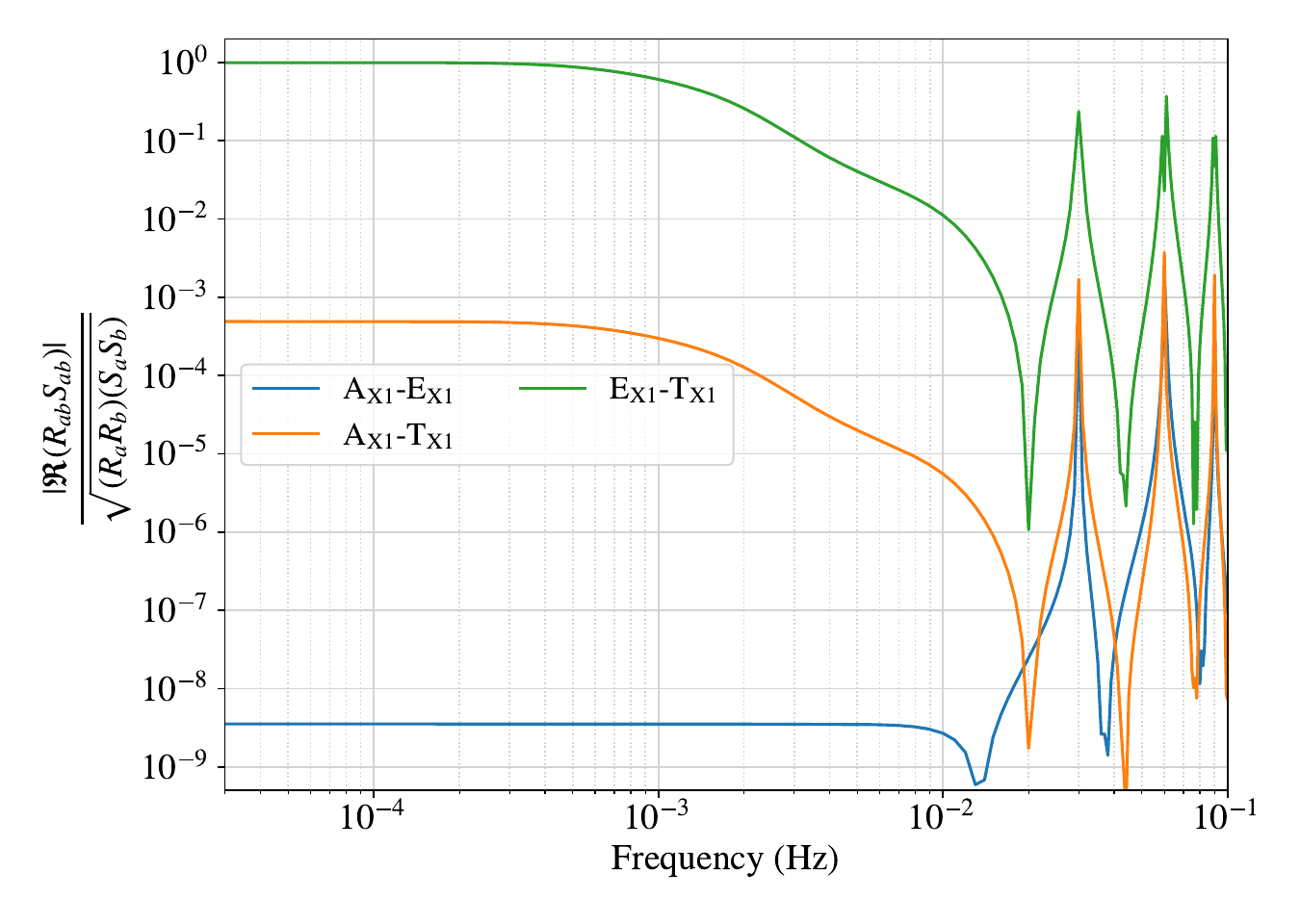}
\includegraphics[width=0.48\textwidth]{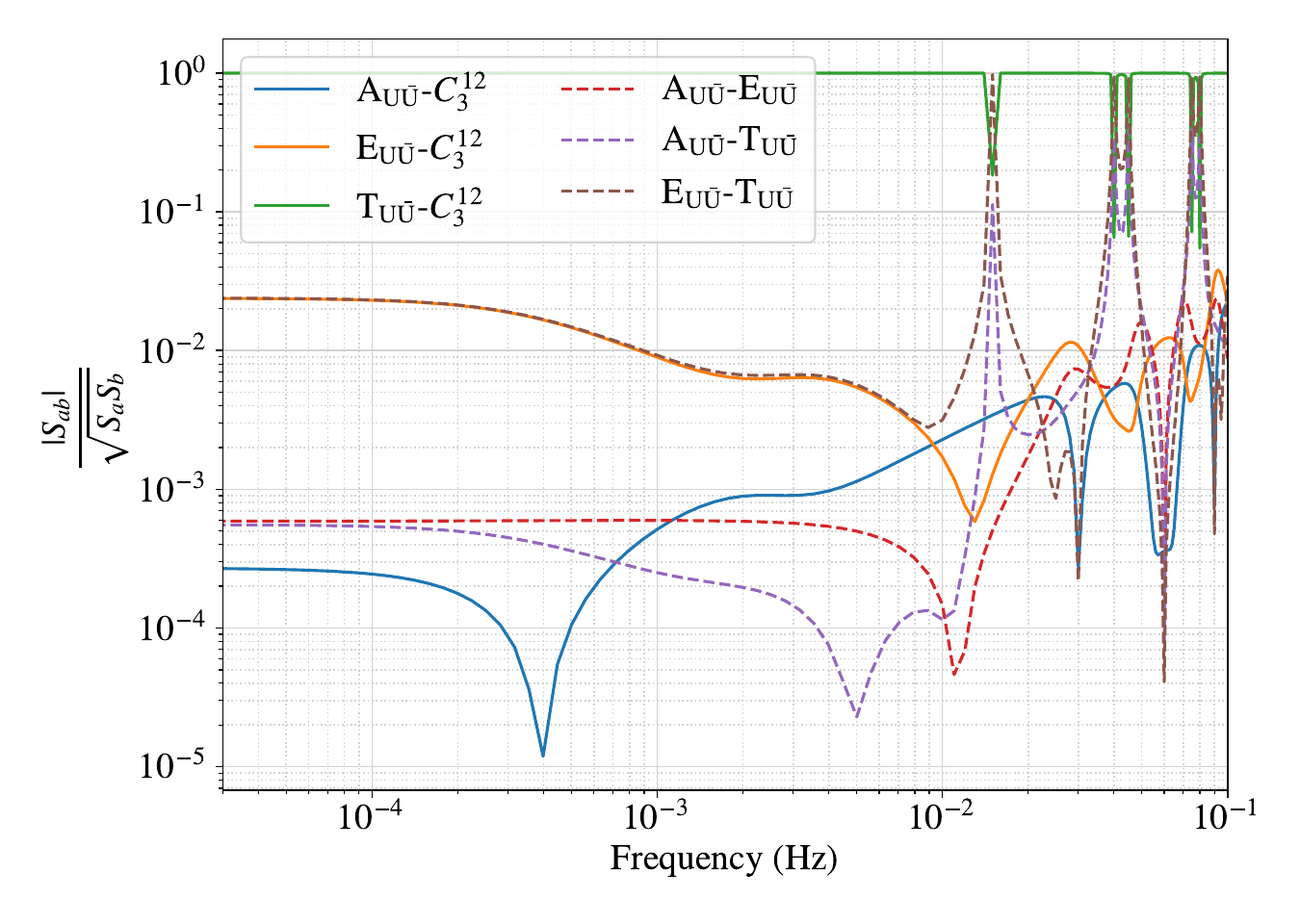} 
\includegraphics[width=0.48\textwidth]{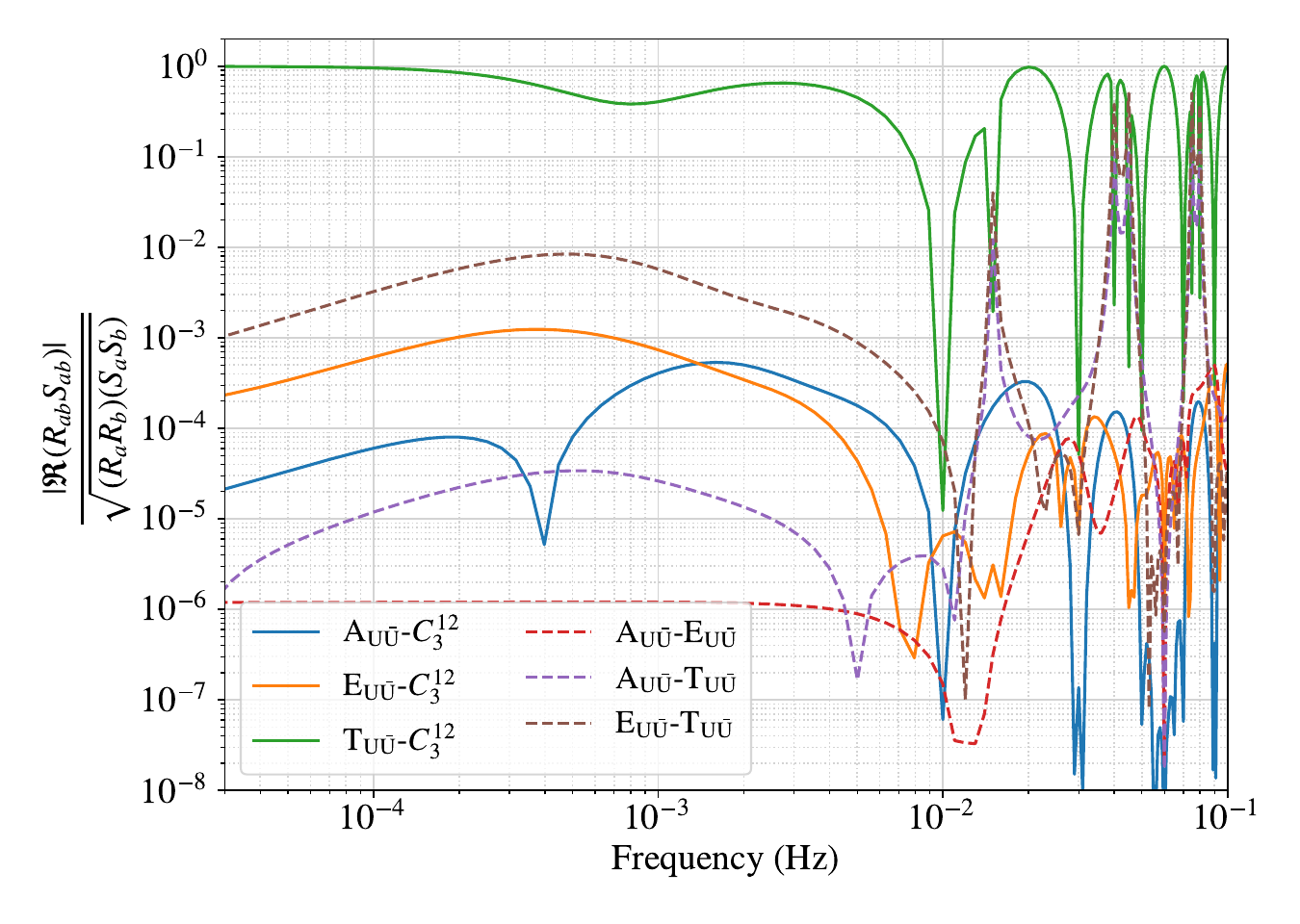}
\caption{\label{fig:cross_noise_response} The cross-correlations between noises and GW responses among optimal TDI channels for Michelson and hybrid Relay. The upper panel depicts correlations for Michelson, while the lower panel shows those for hybrid Relay with additional null stream $C^{12}_3$. The left and right columns display noise and GW response correlations, respectively. In the upper panel, T$_\mathrm{X1}$ is fully correlated with E$_\mathrm{X1}$ at frequencies lower than 1 mHz. The correlations across three optimal channels of hybrid Relay are lower than those of Michelson. T$_\mathrm{U\overline{U}}$ shows relative independence from A$_\mathrm{U\overline{U}}$/E$_\mathrm{U\overline{U}}$ at frequencies lower than 15 mHz. While T$_\mathrm{U\overline{U}}$ and $C^{12}_3$ are highly correlated except at particular null frequencies, $C^{12}_3$ remains independent from A$_\mathrm{U\overline{U}}$/E$_\mathrm{U\overline{U}}$ except around their common null frequencies. }
\end{figure*}

Triple optimal observables from a TDI configuration are orthogonal in a static and equal-arm triangular constellation. However, in reality, the arm lengths between the three S/C will be dynamically unequal during the detector motion. 
In this section, we examine cross-correlation between the TDI observables and the stabilities of spectra for Michelson, hybrid Relay and Sagnac configurations.

The second-generation TDI observables can effectively suppress laser frequency noises \cite[and references therein]{Vallisneri:2004bn,Otto:2015,Bayle:2018hnm,Wang:2020pkk,Muratore:2020mdf,Staab:2023qrb} and mitigate clock noise \cite{Hartwig:2020tdu}.
Acceleration noise and OMS noise are expected to be dominating noises in the data streams, and their respective noise spectra could be \cite{2017arXiv170200786A},
\begin{equation} \label{eq:noise_budgets}
\begin{aligned}
& \sqrt{ \mathrm{ S_{acc} } } = A_\mathrm{acc} \frac{\rm fm/s^2}{\sqrt{\rm Hz}} \sqrt{1 + \left(\frac{0.4 {\rm mHz}}{f} \right)^2 }  \sqrt{1 + \left(\frac{f}{8 {\rm mHz}} \right)^4 }, \\
& \sqrt{ \mathrm{ S_{oms} } } = A_\mathrm{oms} \frac{\rm pm}{\sqrt{\rm Hz}} \sqrt{1 + \left(\frac{2 {\rm mHz}}{f} \right)^4 },
 \end{aligned}
\end{equation}
where $A_\mathrm{acc}$ and $A_\mathrm{oms}$ represent the amplitudes of acceleration noise and OMS noise, respectively. Assuming identical noise characteristics, the amplitudes are set to be $A_\mathrm{acc} = 3$ and $A_\mathrm{oms} = 10$ for all measurement system on each S/C.

The correlations between TDI observables from Michelson and hybrid Relay configurations are depicted in Fig. \ref{fig:cross_noise_response} under a static unequal arm constellation. The upper two plots show that both the noise and GW response of T$_\mathrm{X1}$ are fully correlated with E$_\mathrm{X1}$ for frequencies lower than $\sim$1 mHz. The correlation between A$_\mathrm{X1}$ and T$_\mathrm{X1}$ is relative lower, while A$_\mathrm{X1}$ and E$_\mathrm{X1}$ exhibit much less correlation. 
In the case of Michelson, correlations between the data streams increase around null frequencies $i/(4L)$ Hz ($i=1,2,3...$), which also include the numerical errors due to small values of noise spectra or GW response. 

Conversely, for hybrid Relay, three data streams demonstrate greater independence from each other, as indicated by the dashed lines in lower plots. The highest correlation between E$_\mathrm{U\overline{U}}$ and T$_\mathrm{U\overline{U}}$ is less than 0.03 in the frequency band below 10 mHz. T$_\mathrm{U\overline{U}}$ exhibits high correlation with A$_\mathrm{U\overline{U}}$/E$_\mathrm{U\overline{U}}$ at its characteristic frequencies, which can be as low as 15 mHz. The correlations between $C^{12}_3$ and hybrid Relay observables are depicted by solid lines in the lower two plots. As shown by the green lines, these two data steams are highly correlated except at a few particular frequencies. $C^{12}_3$ has low relevance with A$_\mathrm{U\overline{U}}$/E$_\mathrm{U\overline{U}}$, and their correlations only increase at frequencies $i/(L)$ Hz. The correlations could vary if the noises are non-identical, we evaluated such a case in Appendix \ref{sec:correlation-non-identical}.

\begin{figure*}[thb]
\includegraphics[width=0.48\textwidth]{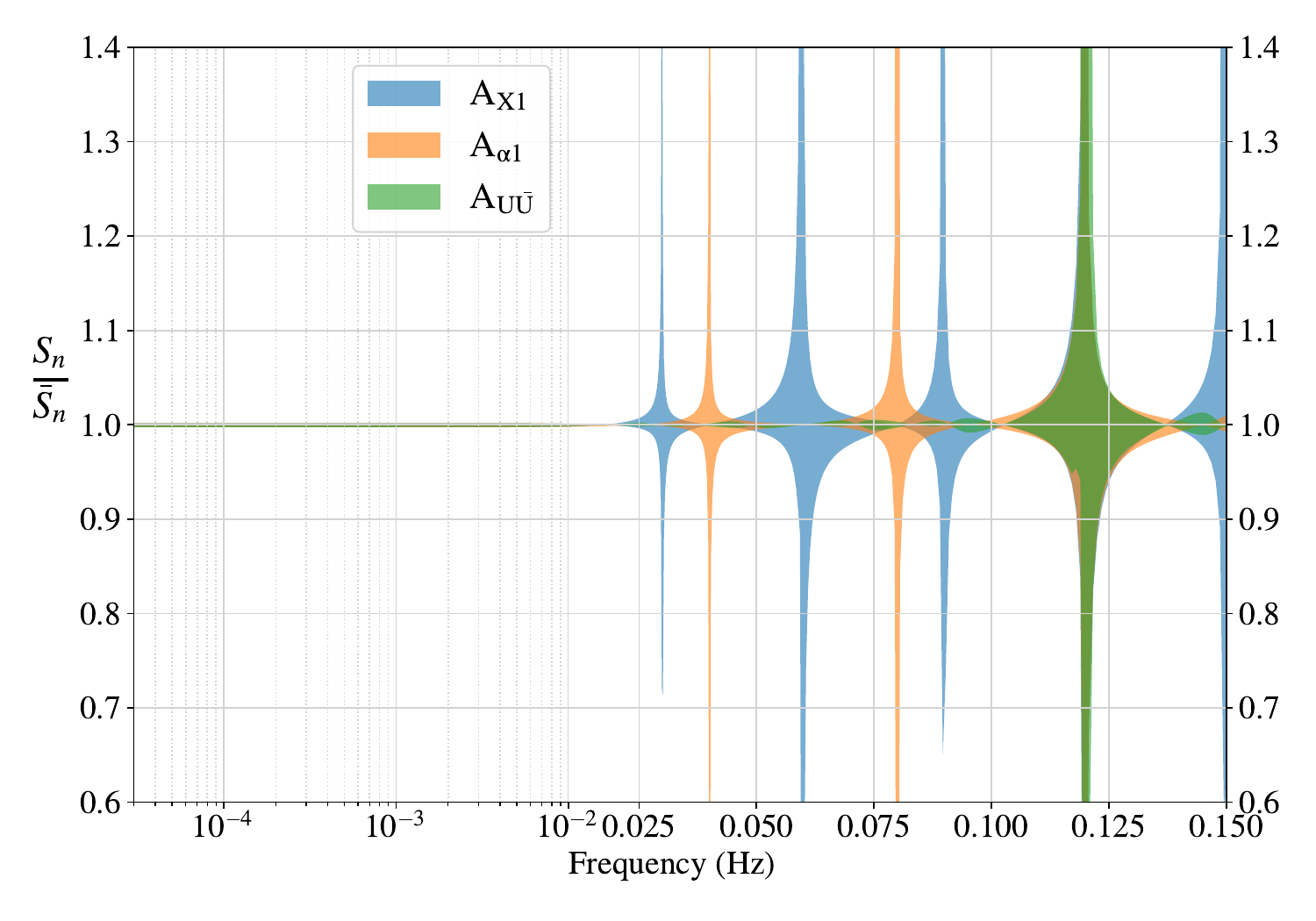}
\includegraphics[width=0.48\textwidth]{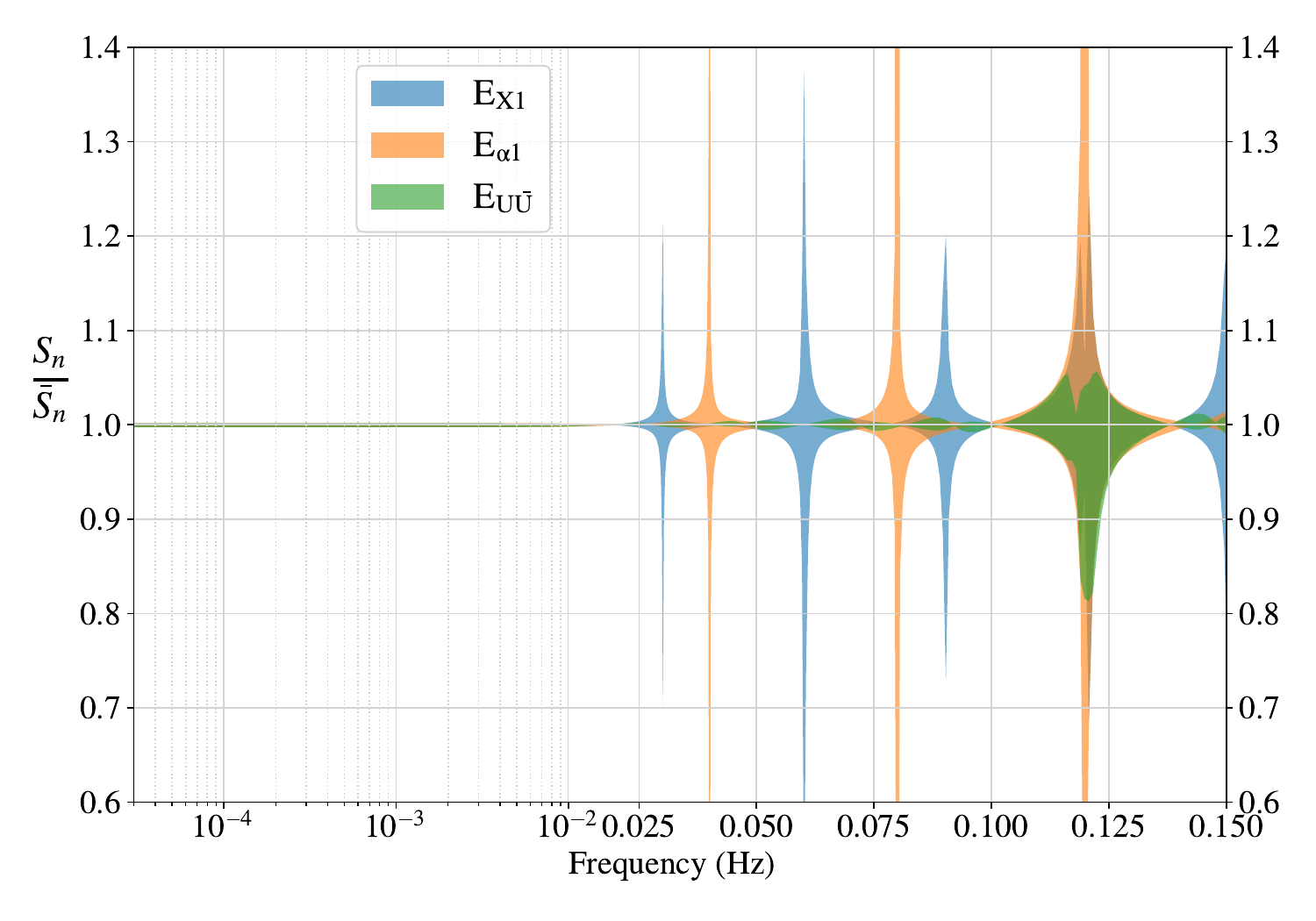}
\includegraphics[width=0.48\textwidth]{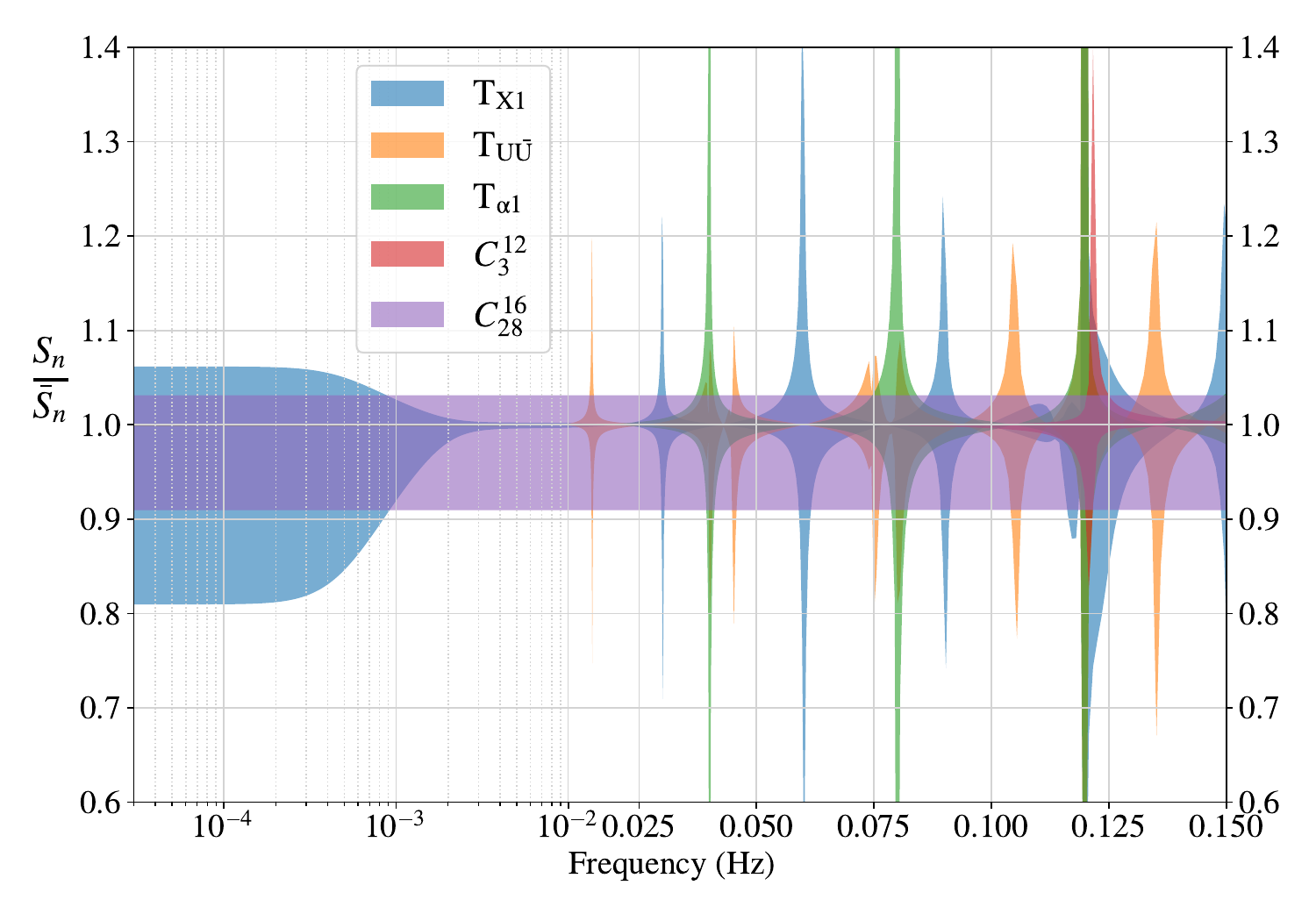}
\includegraphics[width=0.48\textwidth]{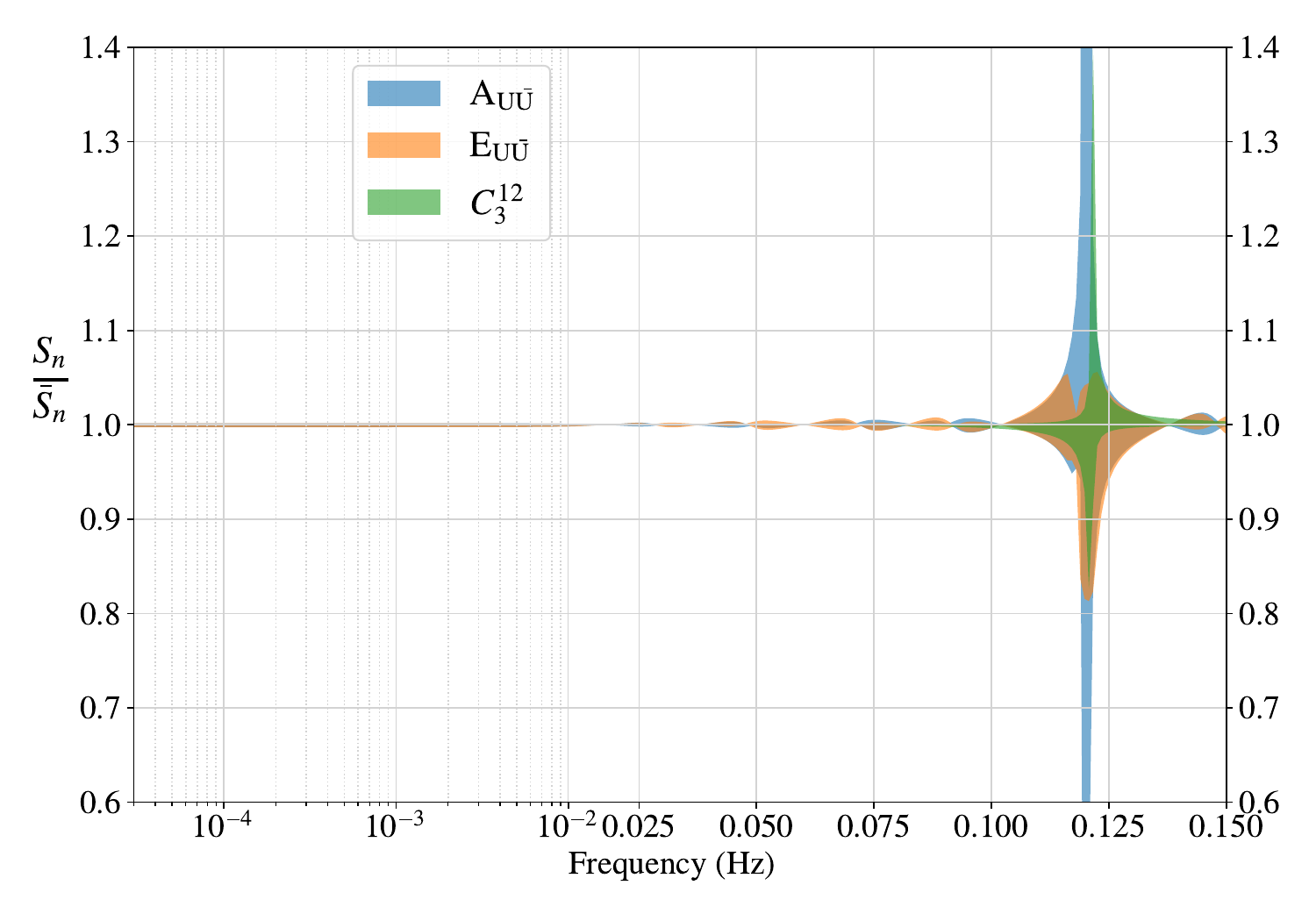}
\caption{\label{fig:uncertainty_AET} Ratios between time-varying PSDs and their averages for optimal TDI channels of Michelson, Sagnac, and hybrid Relay over 30 days, as well as the null streams $C^{12}_3$ and $C^{16}_{28}$. The spectra vary significantly around their null/characteristic frequencies. The characteristic frequencies of Michelson and Sagnac are $\frac{i}{4 L} \simeq 0.03 i$ Hz $(i=1, 2, 3...)$ and $\frac{i}{ 3 L} \simeq 0.04 i$ Hz for LISA mission, respectively. The null frequencies of A/E from hybrid Relay are $\frac{i}{L} \simeq 0.12 i$ Hz, with additional characteristic frequencies appearing in the T$_\mathrm{U\overline{U}}$ which the lowest is $0.015$ Hz as shown in lower panel. For the T$_\mathrm{X1}$, besides its spectrum is unstable at null frequencies, it also exhibits large variance at frequencies lower than $\sim$2 mHz. Among the null streams shown in the lower left plot, $C^{12}_3$ is the most stable with fluctuations only around $\frac{i}{L}$, which is the same as the A$_\mathrm{U\overline{U}}$ and E$_\mathrm{U\overline{U}}$. Therefore, the dataset, (A$_\mathrm{U\overline{U}}$, E$_\mathrm{U\overline{U}}$, $C^{12}_3$), is an optimal combination with the most stable spectra in the targeting frequencies band, as shown in the lower right plane. }
\end{figure*}

The stability of noise spectrum for a TDI channel is more crucial for data analysis. Due to orbital motion, the arm lengths will change dynamically, causing the PSD of a TDI to vary even when assuming stationary instrumental noises. To evaluate spectral variations in dynamic scenario, we utilize a set of numerical orbit from \cite{Wang:2017aqq}, which meet the LISA's requirements with arm length differences less than 1\% and Doppler velocities under 6 m/s \footnote{{https://github.com/gw4gw/LISA-Like-Orbit}}. 

We randomly selected one month to evaluate the PSDs of (A, E, T) channels at different time points for Michelson, Sagnac, and hybrid Relay, as well as the null streams $C^{12}_3$ and $C^{16}_{28}$. The ratios between the spectra and the averaged spectra are calculated, and the results are shown in Fig. \ref{fig:uncertainty_AET}. The upper two plots depict the deviations of two science channels for three TDI configurations, with their PSDs deviating from the average spectra around their respective null frequencies. 
Compared to Michelson observables, which are subject to the null frequencies at $i/(4L)$ Hz, the hybrid Relay only have null frequencies at $1/L$ Hz, and Sagnac observables have null frequencies at $i/(3L)$ Hz. Therefore, two science channels of hybrid Relay are the most robust. The ratios for three T channels are shown in the lower left panel of Fig. \ref{fig:uncertainty_AET}. For the T$_\mathrm{X1}$ observable from Michelson, besides its PSD being unstable around its characteristic frequencies, its spectrum also exhibits instability at frequencies lower than $\sim$2 mHz. The T channel of hybrid Relay has more null frequencies than its A/E channel with the lowest being $1/(8L)$ Hz, and the T$_\mathrm{\alpha1}$ has the unstable noise spectrum around its null frequencies. The PSD ratio of $C^{16}_{28}$ is change with time-varying arm lengths in all targeting band, while the $C^{12}_3$ only has fluctuation around the $i/L$ and is the most stable one among five null data streams. The noise PSD of $C^{12}_3$ are formulated and plotted in Appendix \ref{subsec:null_stream_psd}.

An optimal data stream combination would maximize the frequency band with robust spectra. Additionally, the levels of correlation between optimal channels will not affect the analysis result if a covariance matrix is employed to calculate the likelihood, as shown in Eqs. \eqref{eq:likelihood_fn}-\eqref{eq:cov_mat}. Lower correlations between observables would reduce the error of signal-to-noise ratio calculation when the off-diagonal terms are ignored. 
Based on evaluations of spectral robustness and channel correlation, the data streams, (A$_\mathrm{U\overline{U}}$, E$_\mathrm{U\overline{U}}$, $C^{12}_3$) could be an optimal combination for the data analysis as shown in the lower right plot of Fig.\ref{fig:uncertainty_AET}. In the next section, noise characterization comparisons will be performed for different data combinations.

\section{Noise characterizations} \label{sec:simulation_characterization}

The capabilities for noise characterization are assessed by using mock data. Acceleration noises and OMS noises are generated under assumption of being Gaussian and stationary. Data streams of TDI observables are simulated in the time-domain by using \textsf{SATDI} \cite{Wang:2024ssp}. The sampling rate is set to be 4 Hz, with interpolation implemented during the TDI process \cite{Shaddock:2004ua,Bayle:2019dfu}. The data duration ranges from 30 to 180 days in a numerical mission orbit without considering data gap. After obtaining three ordinary TDI data streams, their (quasi-)orthogonal data streams are generated by applying transformation of Eq. \eqref{eq:abc2AET}. And the transformation would not change the noise characterizations performance, as verified in Appendix \ref{sec:noisechar_optimal_ordinary}.

Using the simulated time-domain data, the amplitudes of acceleration noises and OMS noises are inferred using Bayesian algorithm. The likelihood function is formulated as \cite{Romano:2016dpx}
\begin{equation} \label{eq:likelihood_fn}
\begin{split}
\ln \mathcal{L} (d|\vec{\theta} ) = \sum_{f_i} & \left[  
-\frac{1}{2} \tilde{\mathbf{d}}^T \mathbf{C}^{-1}  \tilde{\mathbf{d}}^\ast   - \ln \left( \det{ 2 \pi \mathbf{C} } \right)
   \right],
\end{split}
\end{equation}
where $\tilde{\mathbf{d}}$ is frequency-domain TDI data vector obtained via Fourier transform. The matrix $\mathbf{C}$ is the correlation matrix of noises from three optimal channels,
\begin{equation} \label{eq:cov_mat}
\begin{aligned}
 \mathbf{C} = & \frac{T_\mathrm{obs}}{4}
 \begin{bmatrix}
S_\mathrm{AA} & S_\mathrm{AE} & S_\mathrm{AT} \\
S_\mathrm{EA} & S_\mathrm{EE} & S_\mathrm{ET} \\
S_\mathrm{TA} & S_\mathrm{TE} & S_\mathrm{TT}
\end{bmatrix},
\end{aligned}
\end{equation}
where $T_\mathrm{obs}$ is the duration of data used for parameter inference. The priors are set to be unformed within the range [0, 40] for the square of acceleration noise amplitude $A^2_\mathrm{acc}$, and [50, 200] for square of OMS noises amplitude $A^2_\mathrm{oms}$. The estimations are performed using the nested sampler \textsf{MultiNest} \cite{Feroz:2008xx,Buchner:2014nha}.

\begin{table*}[tbh]
\caption{\label{tab:posterior}
Checklist for parameter inference for various data configurations. All noise characterizations set a low-frequency cutoff of $3 \times 10^{-5}$ Hz. The checkmarks (\checkmark) indicate that injected values fall within inferred $3\sigma$ credible regions, and crossmarks (\ding{55}) indicate that true value(s) lie outside of the inferred $3\sigma$ credible regions. The double checkmarks show the optimal data combination. The inferred values of $A^2_\mathrm{acc, 12}+A^2_\mathrm{acc, 21}$ and $A^2_\mathrm{oms, 12}$ are selected to represent uncertainties of the parameters.}
\begin{ruledtabular}
{\renewcommand{\arraystretch}{1.2}
\begin{tabular}{cccccccc}
TDI data streams & orbit & $f_\mathrm{max}$ (Hz) & duration (day) & $A^2_\mathrm{acc, 12}+A^2_\mathrm{acc, 21}$ & $A^2_\mathrm{oms, 12}$ & status & plot  \\
\hline
(A$_\mathrm{U\overline{U}}$, E$_\mathrm{U\overline{U}}$, T$_\mathrm{U\overline{U}}$)  & static & $0.01$ &  30 & $ 17.58_{-1.00}^{1.06} $ & $ 107.64^{9.55}_{-10.30} $ & \checkmark & Fig. \ref{fig:corner_static_30days}  \\
(A$_\mathrm{U\overline{U}}$, E$_\mathrm{U\overline{U}}$, T$_\mathrm{U\overline{U}}$) & static & $0.1$ & 30 & $ 17.70_{-0.99}^{1.01} $ & $ 100.19^{1.37}_{-1.33} $ & \checkmark & Fig. \ref{fig:corner_static_30days}  \\
(A$_\mathrm{X1}$, E$_\mathrm{X1}$, T$_\mathrm{X1}$)                             & static & $0.01$ & 30 & $ 17.57_{-0.98}^{1.02} $ & $ 107.49^{10.31}_{-10.37} $ & \checkmark & Fig. \ref{fig:corner_static_30days}   \\
(A$_\mathrm{X1}$, E$_\mathrm{X1}$, T$_\mathrm{X1}$)                              & static & $0.1$ & 30 & $ 17.69_{-0.97}^{1.02} $ & $ 100.21^{1.50}_{-1.40} $ & \checkmark & Fig. \ref{fig:corner_static_30days}  \\
\hline
(A$_\mathrm{U\overline{U}}$, E$_\mathrm{U\overline{U}}$, T$_\mathrm{U\overline{U}}$) & dynamic & $0.01$ & 30 & $ 17.56_{-1.00}^{1.04} $ & $ 107.54^{10.84}_{-10.77} $ & \checkmark & Fig. \ref{fig:corner_dynamic_0p01Hz} \\
(A$_\mathrm{U\overline{U}}$, E$_\mathrm{U\overline{U}}$, T$_\mathrm{U\overline{U}}$) & dynamic & $0.01$ & 90 & $ 17.87_{-0.59}^{0.60} $ & $ 103.55^{5.73}_{-6.26}  $ & \checkmark & Fig. \ref{fig:corner_dynamic_0p01Hz} \\
(A$_\mathrm{U\overline{U}}$, E$_\mathrm{U\overline{U}}$, T$_\mathrm{U\overline{U}}$) & dynamic & $0.01$ & 180 & $ 18.12_{-0.42}^{0.42} $ & $ 96.76^{4.10}_{-4.30}  $ &  \checkmark &  \\
(A$_\mathrm{U\overline{U}}$, E$_\mathrm{U\overline{U}}$, T$_\mathrm{U\overline{U}}$) & dynamic & $0.03$ & 120 & $ 17.18_{-0.47}^{0.47} $ & $ 107.28_{-1.67}^{1.70} $ & \ding{55} & Fig. \ref{fig:corner_hybrid_Relay_ordinary_optimal}  \\
($\mathrm{U\overline{U}}$, $\mathrm{V\overline{V}}$, $\mathrm{W\overline{W}}$) & dynamic & $0.03$ & 120 & $ 17.17_{-0.48}^{0.49} $ & $ 107.29_{-1.65}^{1.68} $ & \ding{55} & Fig. \ref{fig:corner_hybrid_Relay_ordinary_optimal} \\
(A$_\mathrm{U\overline{U}}$, E$_\mathrm{U\overline{U}}$, T$_\mathrm{U\overline{U}}$) & dynamic & $0.1$ & 30 & $ 17.83_{-0.98}^{1.01} $ & $ 105.24^{1.52}_{-1.53} $ & \ding{55} &  \\
(A$_\mathrm{X1}$, E$_\mathrm{X1}$, T$_\mathrm{X1}$)                             & dynamic & $0.01$ & 30 & $ 15.20_{-0.94}^{0.96} $ & $ 161.62^{21.48}_{-22.42} $ &  \ding{55} & Fig. \ref{fig:corner_dynamic_0p01Hz} \\
(A$_\mathrm{X1}$, E$_\mathrm{X1}$, T$_\mathrm{X1}$)                             & dynamic & $0.01$ & 90 & $ 28.83_{-0.08}^{0.16} $ & $ 194.78^{0.05}_{-0.04} $ &  \ding{55} & Fig. \ref{fig:corner_dynamic_0p01Hz} \\
(A$_\mathrm{X1}$, E$_\mathrm{X1}$, T$_\mathrm{X1}$)                             & dynamic & $0.1$ & 30 & $ 17.24_{-0.92}^{0.92} $ & $ 102.26^{1.54}_{-1.57} $ & \ding{55} &   \\
\hline
(A$_\mathrm{U\overline{U}}$, E$_\mathrm{U\overline{U}}$) & static & $0.1$ & 90 & $ 18.25_{-0.57}^{0.60} $  & $ 108.14^{33.11}_{-35.49} $ &  \checkmark & Fig. \ref{fig:corner_AE_0p1Hz} \\
(A$_\mathrm{X1}$, E$_\mathrm{X1}$)                    & static & $0.1$ & 90 & $ 18.25_{-0.56}^{0.57} $ & $ 96.56^{15.17}_{-13.54} $ & \checkmark & Fig. \ref{fig:corner_AE_0p1Hz} \\
(A$_\mathrm{U\overline{U}}$, E$_\mathrm{U\overline{U}}$) & dynamic & $0.1$ & 90 & $ 17.98_{-0.58}^{0.59} $ & $ 94.46_{-44.10}^{43.90} $ &  \checkmark & Fig. \ref{fig:corner_AE_0p1Hz} \\
(A$_\mathrm{X1}$, E$_\mathrm{X1}$)                    & dynamic & $0.1$ & 90 & $ 17.97_{-0.56}^{0.57} $  & $ 106.72_{-19.62}^{18.58} $ & \checkmark & Fig. \ref{fig:corner_AE_0p1Hz} \\
\hline
(A$_\mathrm{U\overline{U}}$, E$_\mathrm{U\overline{U}}$, T$_\mathrm{\alpha1}$) & dynamic & $0.03$ & 120 & $ 17.98_{-0.49}^{0.49} $ &  $ 99.51^{1.44}_{-1.48} $ &  \checkmark &    \\
(A$_\mathrm{U\overline{U}}$, E$_\mathrm{U\overline{U}}$, T$_\mathrm{\alpha1}$) & dynamic & $0.03$ & 180 & $ 18.12_{-0.41}^{0.41} $ & $ 99.17^{1.23}_{-1.17} $ &  \ding{55} &  \\
(A$_\mathrm{U\overline{U}}$, E$_\mathrm{U\overline{U}}$, $C^{12}_3$) & dynamic & $0.1$ & 180 & $ 18.12_{-0.42}^{0.41} $ & $ 99.62^{0.57}_{-0.54} $ &   \checkmark  \checkmark & Fig. \ref{fig:corner_dynamic_AE_zeta12_0p1Hz} 
\end{tabular}
}
\end{ruledtabular}
\end{table*}

Two scenarios are implemented for noise characterizations: 1) simulating data and inferring parameter assuming a static unequal-arm constellation (where arm lengths remain constant over time), and 2) generating data and estimating parameters for a dynamic constellation (where arm lengths vary with detector's orbit). The first case aims to calibrate noise characterization algorithm and verify the capabilities of different TDI configurations under conditions where arm length do not change. The second scenario evaluates the impact of dynamic arm lengths on the noise characterization compared to the static case, revealing varying capabilities across different TDI configurations.

During the characterization, two adjustable factors are considered: the effective frequency band and the duration of the noise data. Regarding the frequency band, the low-frequency cutoff is set to be 0.03 mHz, and the high frequency cutoff varies depending on the specific evaluations. The frequency range of [0.03, 10] mHz is considered a 'clean' band, where the correlation between three optimal channels are steady and the spectra remain stable over a month's duration (except the Michelson-T channel), as shown in Figs. \ref{fig:cross_noise_response} and \ref{fig:uncertainty_AET}. The duration of the data also plays a crucial role in determining the amount of data and the frequency resolution. Opting for a long duration increases spectral resolution but may introduce more fluctuations. Therefore, a trade-off is necessary when selecting the data duration and frequency range.
Table \ref{tab:posterior} provides a checklist for parameter inference with various data setups and the feasibility. The inferred values of $A^2_\mathrm{acc, 12}+A^2_\mathrm{acc, 21}$ and $A^2_\mathrm{oms, 12}$ ($A_{ij}$ represents the noise component on S/C$i$ facing S/C$j$) are selected to represent uncertainties of the parameters. Checkmarks and crossmarks indicate whether the inference correctly or incorrectly encompass the input true values.

\begin{figure*}[htb]
\includegraphics[width=0.44\textwidth]{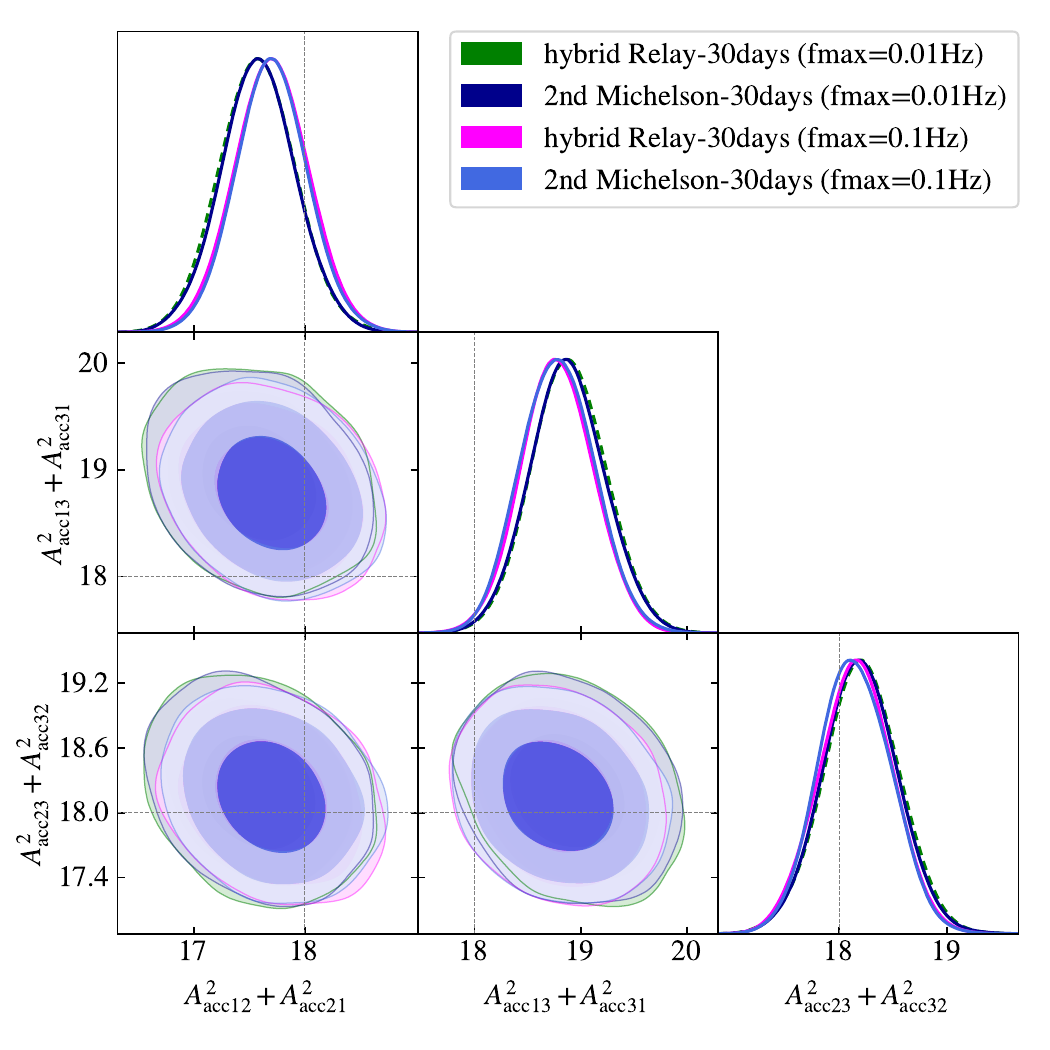}
\includegraphics[width=0.44\textwidth]{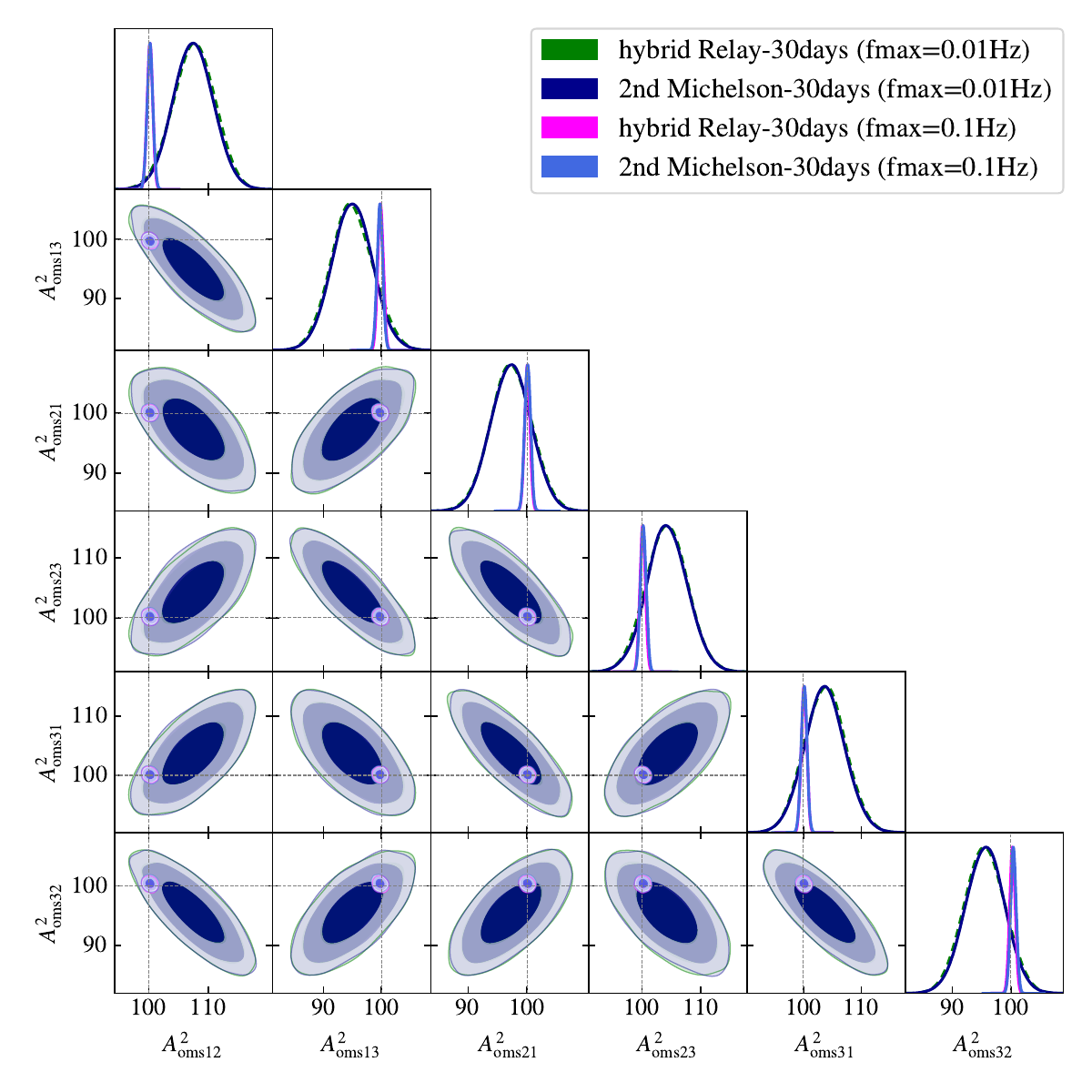}
\caption{\label{fig:corner_static_30days} 
The inferred parameter distributions from 30-day data (A, E, T) for a static unequal-arm constellation with high frequency cutoffs of 0.01 Hz and 0.1 Hz. Acceleration noises dominates the spectra at frequencies lower than $\sim$4 mHz, and increasing high frequency cutoff from 0.01 Hz to 0.1 Hz introduces more OMS noise data rather than acceleration noise. Therefore, the amplitudes of acceleration noise are estimated with same constraint in the left plot, and the constraints on OMS noise are improved with a higher high frequency cutoff in the right plot. Results from Michelson and hybrid Relay are overlapped for the same setup. ($A_{ij}$ represents the noise component on S/C$i$ facing S/C$j$; the three color gradients show the 1$\sigma$ to 3$\sigma$ credible regions).
}
\end{figure*}

The inferred noise parameters for a static unequal-arm constellation with different data setups are depicted in Fig. \ref{fig:corner_static_30days}. This figure illustrates the inferred distributions of noise parameters from 30-day data streams (A, E, T) with a high frequency cutoff of 0.01 Hz or 0.1 Hz.
Due to the degeneracy of two acceleration noises in an interferometric link, the sums of amplitude squares are shown in the left plot. The right panel displays the resolved individual six amplitudes of OMS noises. In the current scenario, Michelson and hybrid Relay exhibit identical performances under same setup, verifying that the different levels of correlation between the optimal data streams do \textit{not} (significantly) affect the inference results. Increasing the high frequency cutoff from 0.01 Hz to 0.1 Hz does not improve the characterization of acceleration noise, as this noise dominates the spectra at frequencies lower than $\sim$4 mHz. Widening the higher frequency band contributes additional data related to OMS noise rather than acceleration noise. Therefore, uncertainties for OMS noises would be significantly reduced with more high-frequency data. It is noted that the null frequencies are excluded to reduce numerical error if them fall within employed frequency band.

\begin{figure*}[hbt]
\includegraphics[width=0.44\textwidth]{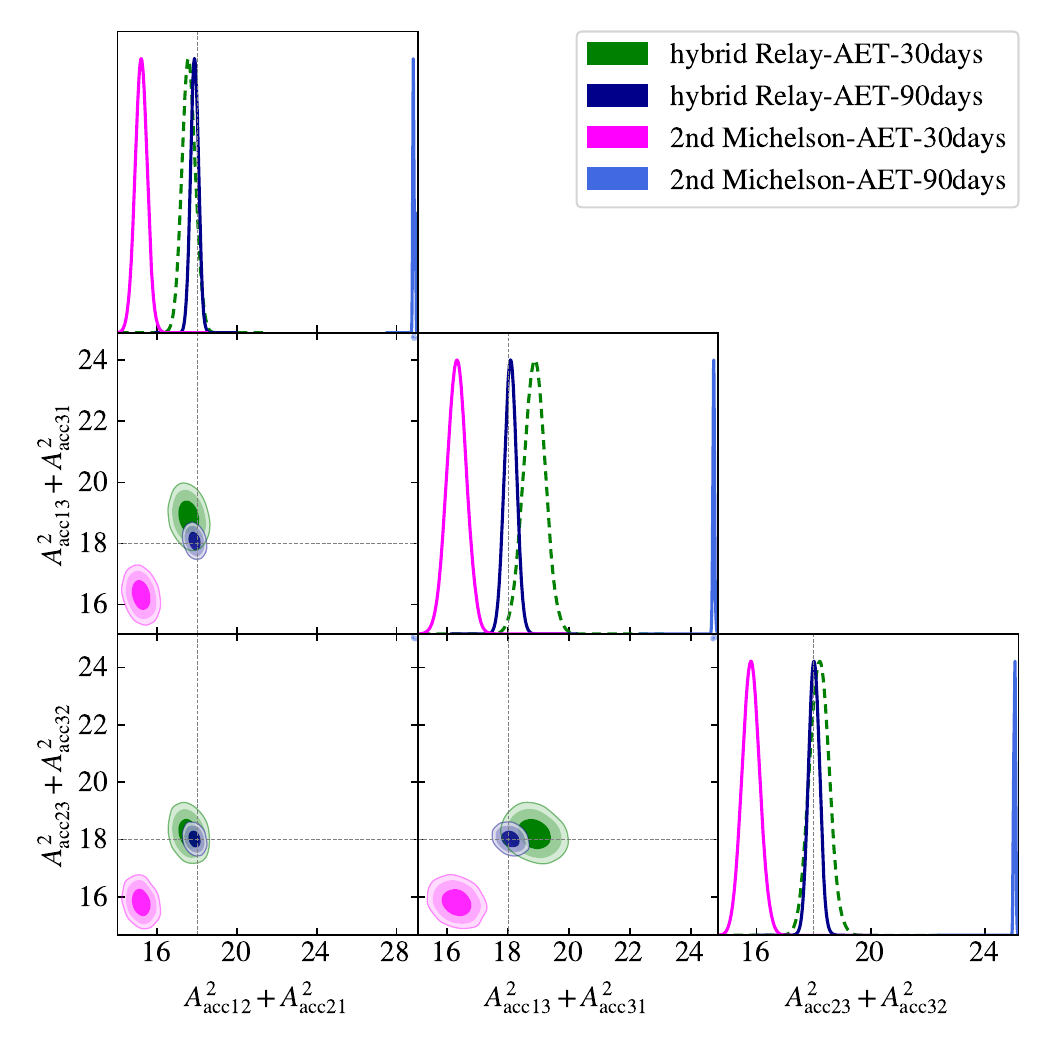}
\includegraphics[width=0.44\textwidth]{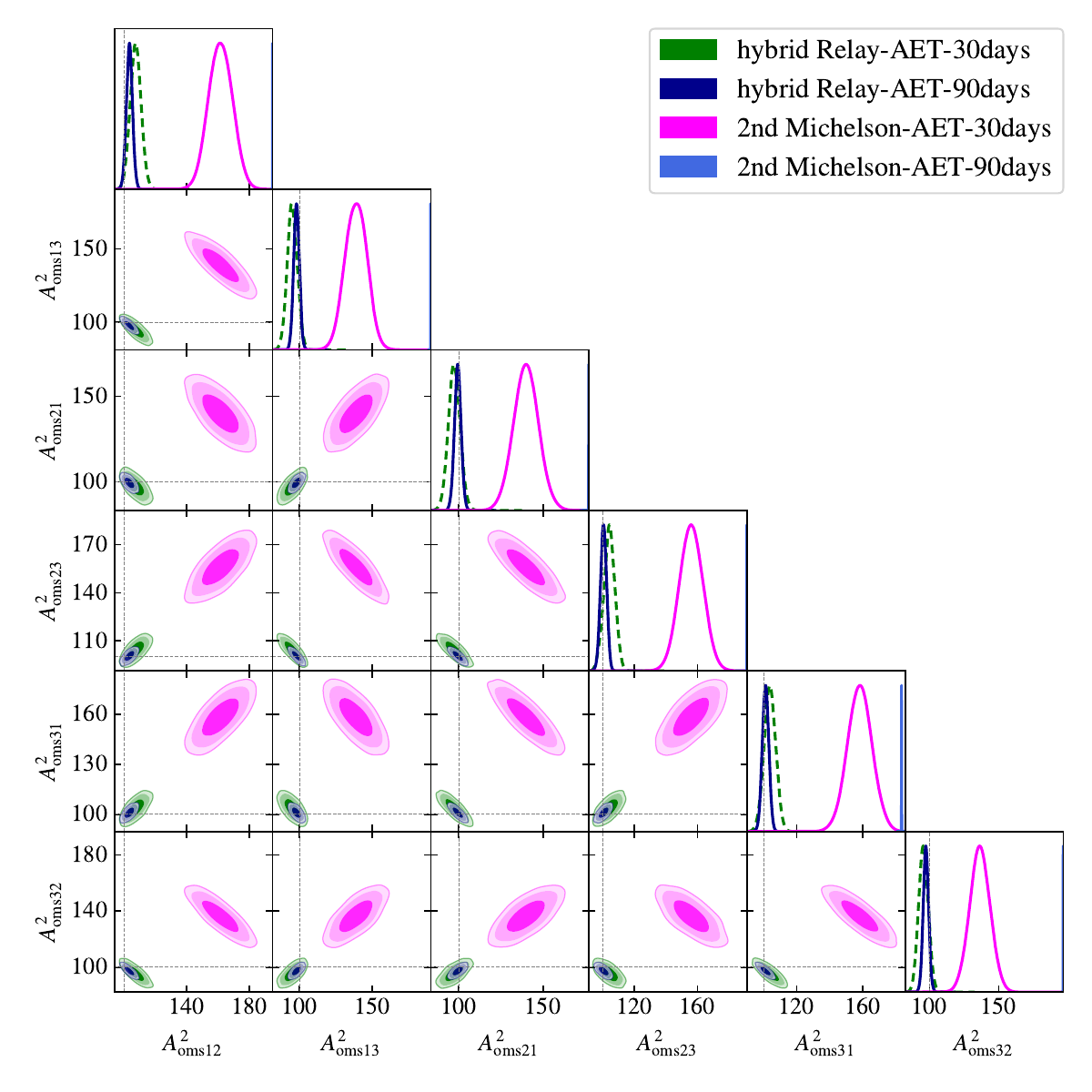}
\caption{\label{fig:corner_dynamic_0p01Hz} The inferred parameter distributions from 30-day and 90-day dataset (A, E, T) based on dynamic constellation motion with a high frequency cutoff of 0.01 Hz. Michelson's results diverge from the true values due to its time-varying PSD/CSD with T channel. In contrast, the inference results from hybrid Relay are more sensible and robust. A longer data duration improves data constraints on both types of noise parameters. }
\end{figure*}

The inferred parameter distributions using three optimal data from the dynamic case are shown in Fig. \ref{fig:corner_dynamic_0p01Hz}. Comparisons between the Michelson and hybrid Relay use data durations of 30 days and 90 days with a high frequency cutoff of 0.01 Hz. Because the noise spectra with T$_\mathrm{X1}$ are unstable and sensitive to the changes in arm lengths, as analyzed in Fig. \ref{fig:uncertainty_AET}, the inferred distributions from Michelson fail to encompass the true values for either acceleration noise or OMS noise. In contrast, the hybrid Relay effectively constrains the parameters within reasonable regions for both types of noises. Extending the data duration from 30 days to 90 days enhances the precision of result from hybrid Relay but worsens the inference from Michelson.

\begin{figure*}[htb]
\includegraphics[width=0.44\textwidth]{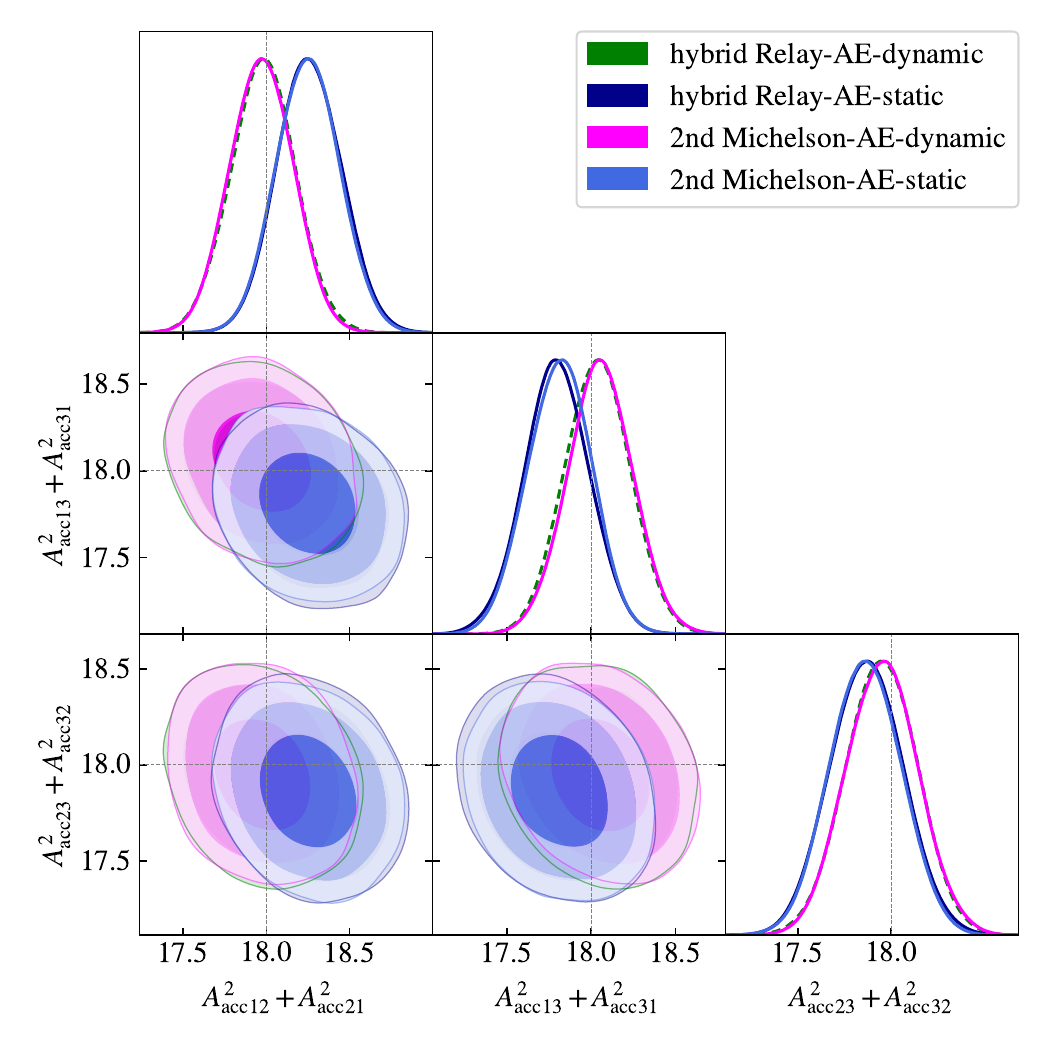}
\includegraphics[width=0.44\textwidth]{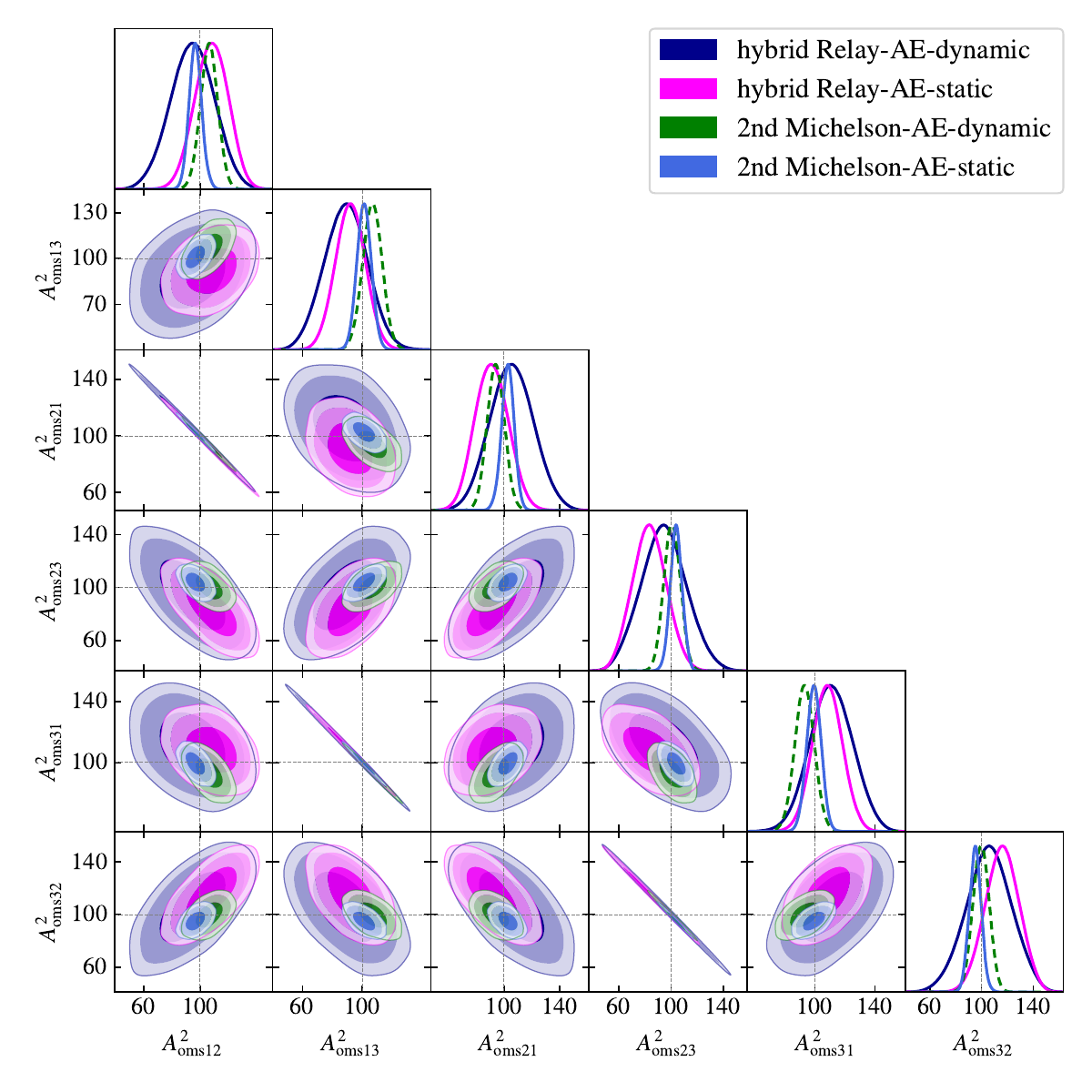}
\caption{\label{fig:corner_AE_0p1Hz} The inferred parameter distributions from 90-day data streams (A, E) based on static and dynamic unequal-arm constellation with a high frequency cutoff 0.1 Hz. The determinations on acceleration noises are identical for Michelson and hybrid Relay with same data setups. In the right plot, Michelson achieves more precise distributions for OMS noise compared to hybrid Relay. The degeneracy between different OMS noise components could not be effective broken using (A, E) compared to results from (A, E, T). Compared to the static case, results from dynamic case exhibit larger uncertainties because the unstable spectra around null frequencies are not fully gated. }
\end{figure*}

If the T channel is excluded, two science data stream can characterize the noise parameters within appropriate ranges, as shown in Fig. \ref{fig:corner_AE_0p1Hz}. Comparisons are made by using 90 days of (A, E) data with a 0.1 Hz high frequency cutoff for both static and dynamic case. As depicted in the left plot, the determinations for acceleration noise are identical for both scenarios. However, Michelson outperforms hybrid Relay in resolving OMS noise parameters, likely due to its relatively better ability to break the degeneracies between the parameters. When comparing the distributions from two orbital scenarios, the dynamic case yields larger uncertainties than the static case, which should be caused by fluctuations of spectra around the null frequencies after gating. 
Compared to the static results that including T channel, using only the A and E channels results in much looser constraints on the values of OMS parameters. This also demonstrates the effectiveness of null observable in charactering noises.

\begin{figure*}[htb]
\includegraphics[width=0.44\textwidth]{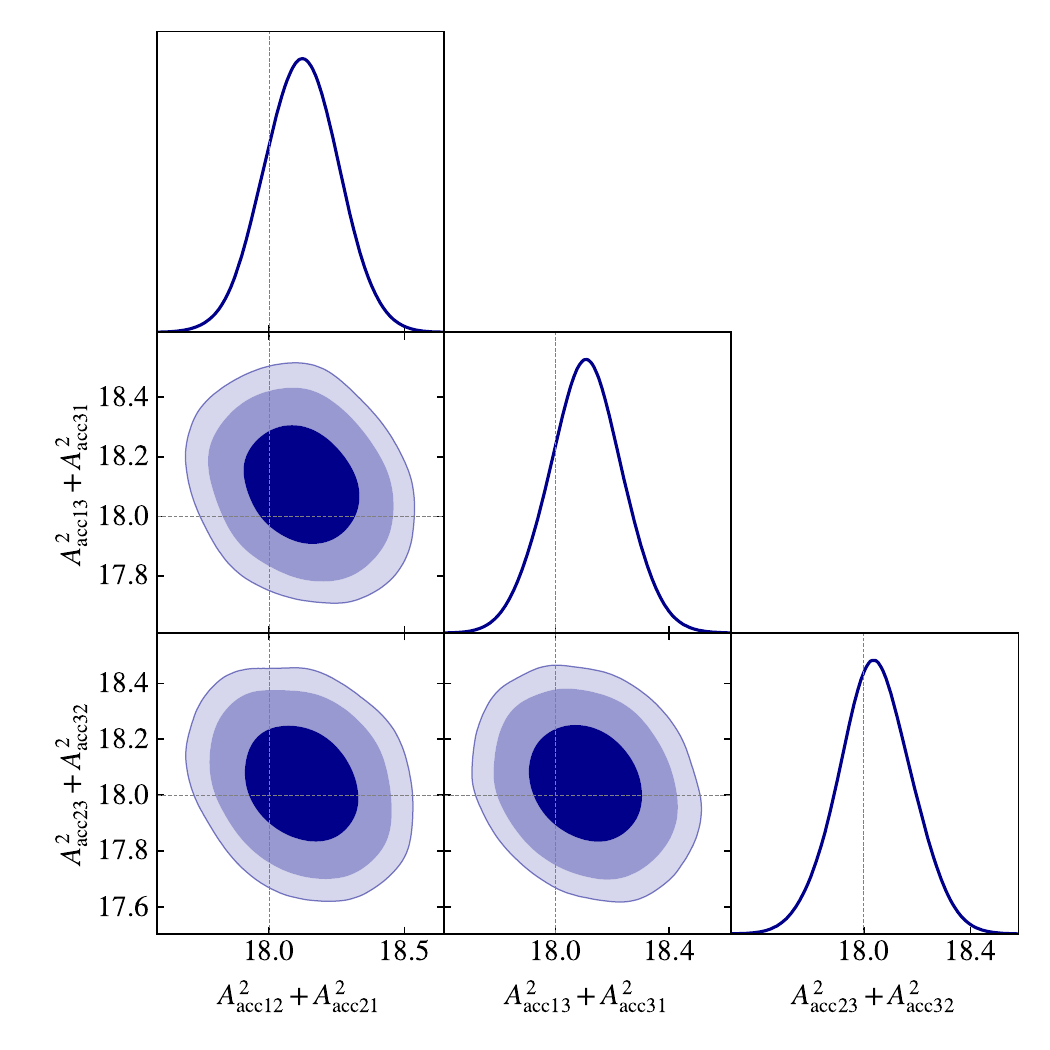}
\includegraphics[width=0.44\textwidth]{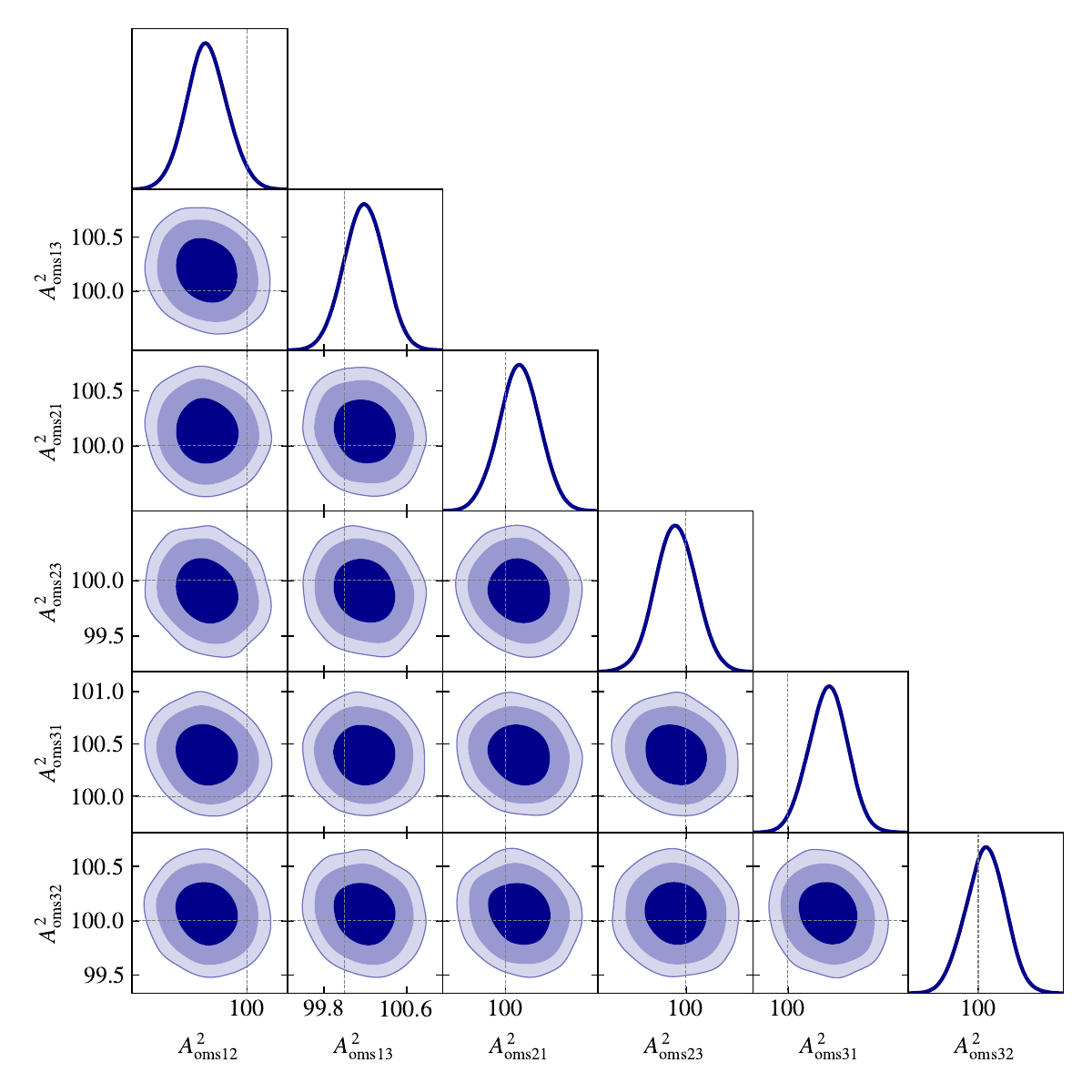}
\caption{\label{fig:corner_dynamic_AE_zeta12_0p1Hz} The inference results from 180-day datasets using $(\mathrm{A_{U\overline{U}}}, \mathrm{E_{U\overline{U}}}, C^{12}_3 )$ in dynamic case with a high frequency cutoff of 0.1 Hz. After substituting $\mathrm{T_{U\overline{U}}}$ with $C^{12}_3$, the new data combination can determine the noise parameters more accurately and precisely compared to the results from other case. }
\end{figure*}

Based on these analysis, the T channel proves to be the 'shortest plank in the barrel' in both Michelson and hybrid Relay. In Michelson configuration, the PSD/CSD involving T$_\mathrm{X1}$ channel fluctuates significantly with mission time in the lower frequency band, undermining Michelson's capability for noise characterizing. On the other hand, in the hybrid Relay configuration, T$_\mathrm{U\overline{U}}$ exhibits more unstable null frequencies than its science channels as illustrated in Fig. \ref{fig:uncertainty_AET}, and its correlations with A$_\mathrm{U\overline{U}}$ and E$_\mathrm{U\overline{U}}$ soar around these null frequencies. T$_\mathrm{\alpha1}$ from Sagnac exhibits more robust noise spectra compared to the T$_\mathrm{U\overline{U}}$, making it a substitute for T$_\mathrm{U\overline{U}}$ given their high correlation. In this case, extending the high frequency cutoff to 30 mHz avoids unstable null frequency for three optimal observables $(\mathrm{A_{U\overline{U}}}, \mathrm{E_{U\overline{U}}}, \mathrm{T_{\alpha1}})$. The inference results from this combination using 120 days with a 0.03 Hz high frequency cutoff can yield sensible parameter distributions, but using 180 days of data fails to accurately infer all 12 parameters.

As the optimal combination, the results of $(\mathrm{A_{U\overline{U}}}, \mathrm{E_{U\overline{U}}}, C^{12}_3)$ dataset using 180 days with 0.1 Hz high frequency cutoff are shown in Fig. \ref{fig:corner_dynamic_AE_zeta12_0p1Hz}. The inferred distributions accurately and precisely encompass the true values compared to previous data combinations, with uncertainties being less than a percents of the true values for OMS noise. Compare to the result from three optimal TDI observables from hybrid Relay with 180-day data, as shown in Table \ref{tab:posterior}, the novel combination significantly reduce the uncertainties of the OMS noise parameters by $\sim$10 times, benefiting from the extended high frequency cutoff from 0.01 Hz to 0.1 Hz. This improvement is much greater compared to the correctly inferred results from the two Michelson science channels, (A$_\mathrm{X1}$, E$_\mathrm{X1}$). The constraints on the parameters of acceleration noise are not improved since the OMS noise dominates frequency band [0.01, 0.1] Hz.

\section{Conclusion and discussion} \label{sec:conclusions}

In this work, we compare the noise characterizations by varying frequency bands and data duration with different data combinations for Michelson and hybrid Relay TDI configurations. Our findings highlight that the hybrid Relay is more effective in determining noise parameters compared to the Michelson configuration. The performance of Michelson is constrained by the unstable spectra with T channel and fluctuations around numerous null frequencies during the orbital evolution of detector. In contrast, the noise spectra of science observables from hybrid Relay are more robust across a wider frequency range due to the reduced occurrence of null frequencies. The efficiency of hybrid Relay can be further enhanced by replacing its T observable with the null stream $C^{12}_3$, where two observables are highly correlated and the latter has as few null frequencies as the science observables of hybrid Relay.

In paper I, we evaluated the performance of two TDI configurations in analyzing chirp signals from massive black hole binary coalescences, revealing deficiencies in the Michelson TDI observables. In that context, the T or null channel plays a negligible role as the analysis was primarily influenced by the two science channel A and E. Our current investigation underscores the critical importance of a reliable null channel in breaking degeneracies between noises and accurately determining noise parameters. 

Our current investigations estimate the amplitudes of noise spectra, which are sufficient to demonstrate the robustness of TDI observables for noise characterization. However, the shape of noise spectra may not be well predicted in real observations, and inferring only the parameters of amplitudes may not suffice in practical scenarios. A more generalized formulation of spectra should be employed to characterize the noise.
On the other hand, it is essential for data analysis to employ a set of data streams with robust spectra. The Michelson TDI configuration is not suitable for this purpose in dynamic orbit scenario. The hybrid Relay proves to be a better choice for the scientific data analysis. We are committed to further examining its performances with different GW sources in future studies.

\begin{acknowledgments}

G.W. acknowledges Olaf Hartwig's valuable comments, which resulted in the the inclusion of the null stream $C^{12}_3$ into the optimal dataset. G.W. also thanks Martina Muratore, Quentin Baghi, Jean-Baptiste Bayle, Shichao Wu for their helpful discussions during the 15th International LISA Symposium 2024 and online discussions.
G.W. was supported by the National Key R\&D Program of China under Grant No. 2021YFC2201903, and NSFC No. 12003059. This work made use of the High Performance Computing Resource in the Core Facility for Advanced Research Computing at Shanghai Astronomical Observatory.
This work are performed by using the python packages $\mathsf{numpy}$ \cite{harris2020array}, $\mathsf{scipy}$ \cite{2020SciPy-NMeth}, $\mathsf{pandas}$ \cite{pandas}, 
$\mathsf{MultiNest}$ \cite{Feroz:2008xx} and $\mathsf{PyMultiNest}$ \cite{Buchner:2014nha}, and the plots are make by utilizing $\mathsf{matplotlib}$ \cite{Hunter:2007ouj}, $\mathsf{GetDist}$ \cite{Lewis:2019xzd}. 

\end{acknowledgments}

\appendix

\section{Correlations between TDI with non-identical noise setup} \label{sec:correlation-non-identical}

\begin{figure*}[htb]
\includegraphics[width=0.48\textwidth]{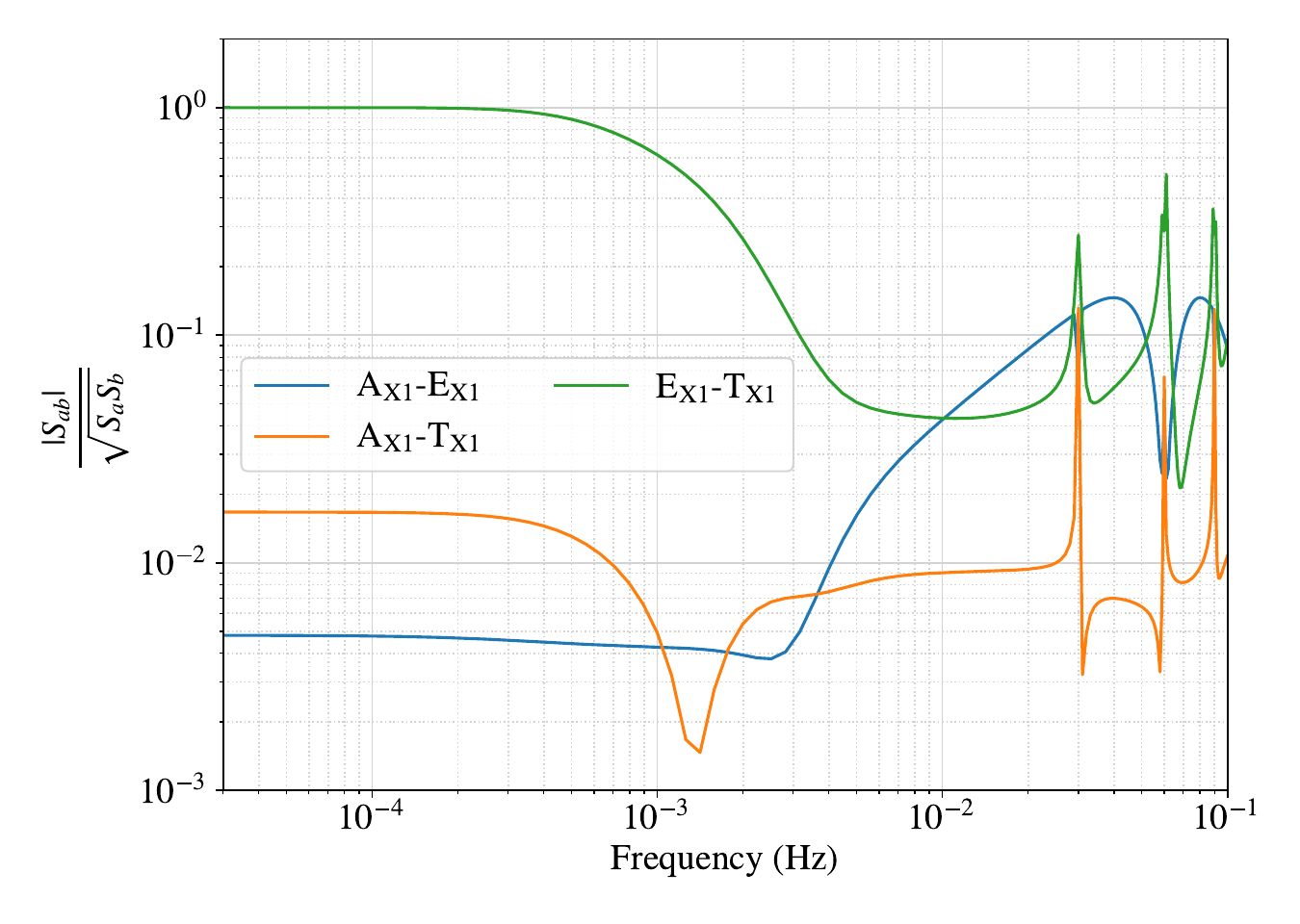}
\includegraphics[width=0.48\textwidth]{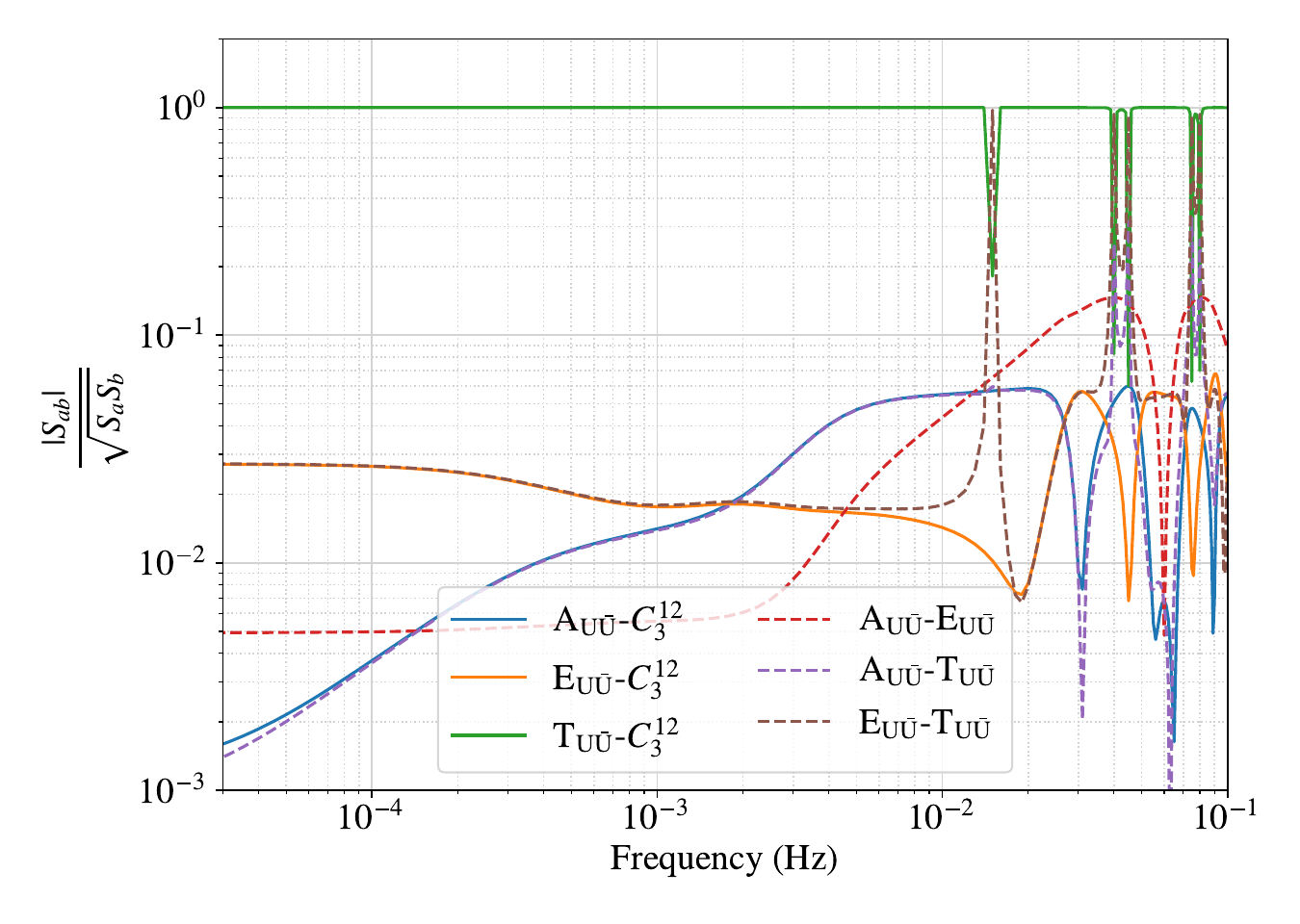}
\caption{\label{fig:correlation_non_identical_noise} The correlations of spectra between optimal TDI channels for Michelson and hybrid Relay with non-identical noises amplitudes. }
\end{figure*}

In current investigation, the amplitudes for acceleration/OMS noises are initially assumed to be identical across different MOSA (Moving Optical Sub-Assemblies). However, these amplitudes could be non-identical and lead to different noise correlations between TDI channels. Assuming a 10\% standard deviation around the fiducial values, the amplitude are randomly sampled and reassigned to the noise budgets:
[$A_\mathrm{oms12}$, $A_\mathrm{oms13}$, $A_\mathrm{oms21}$, $A_\mathrm{oms23}$, $A_\mathrm{oms31}$, $A_\mathrm{oms32}$]=[8.90, 10.14, 11.12, 9.56, 8.74, 10.39],
[$A_\mathrm{acc12}$, $A_\mathrm{acc13}$, $A_\mathrm{acc21}$, $A_\mathrm{acc23}$, $A_\mathrm{acc31}$, $A_\mathrm{acc32}$]=[3.20, 3.03, 2.84, 3.06, 2.75, 3.04]. The correlations between the noise spectra are depicted in Fig. \ref{fig:correlation_non_identical_noise}. Compared to the result in Fig. \ref{fig:cross_noise_response}, the T$_\mathrm{X1}$ and E$_\mathrm{X1}$ remain correlated in low frequencies, and T$_\mathrm{\alpha1}$ remains highly correlated with T$_\mathrm{U\overline{U}}$. The correlations between other channels increase, especially for the A and E observables. As a caveat, to accurately analyze the data, the analysis using covariance matrix Eqs. \eqref{eq:likelihood_fn}-\eqref{eq:cov_mat} formulas should to be employed to account for their cross-correlations.

\section{noise characterization with optimal and ordinary observables} \label{sec:noisechar_optimal_ordinary}

\begin{figure*}[htb]
\includegraphics[width=0.48\textwidth]{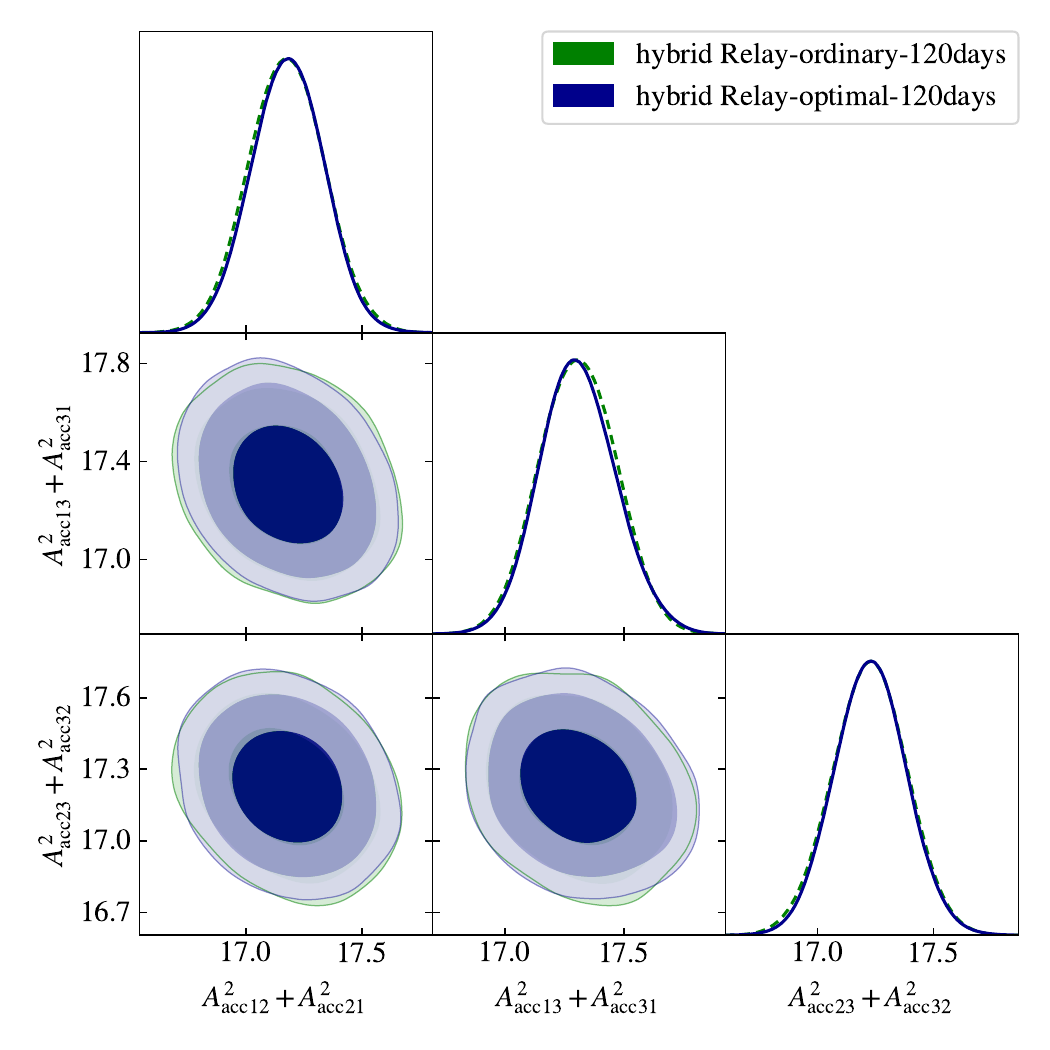}
\includegraphics[width=0.48\textwidth]{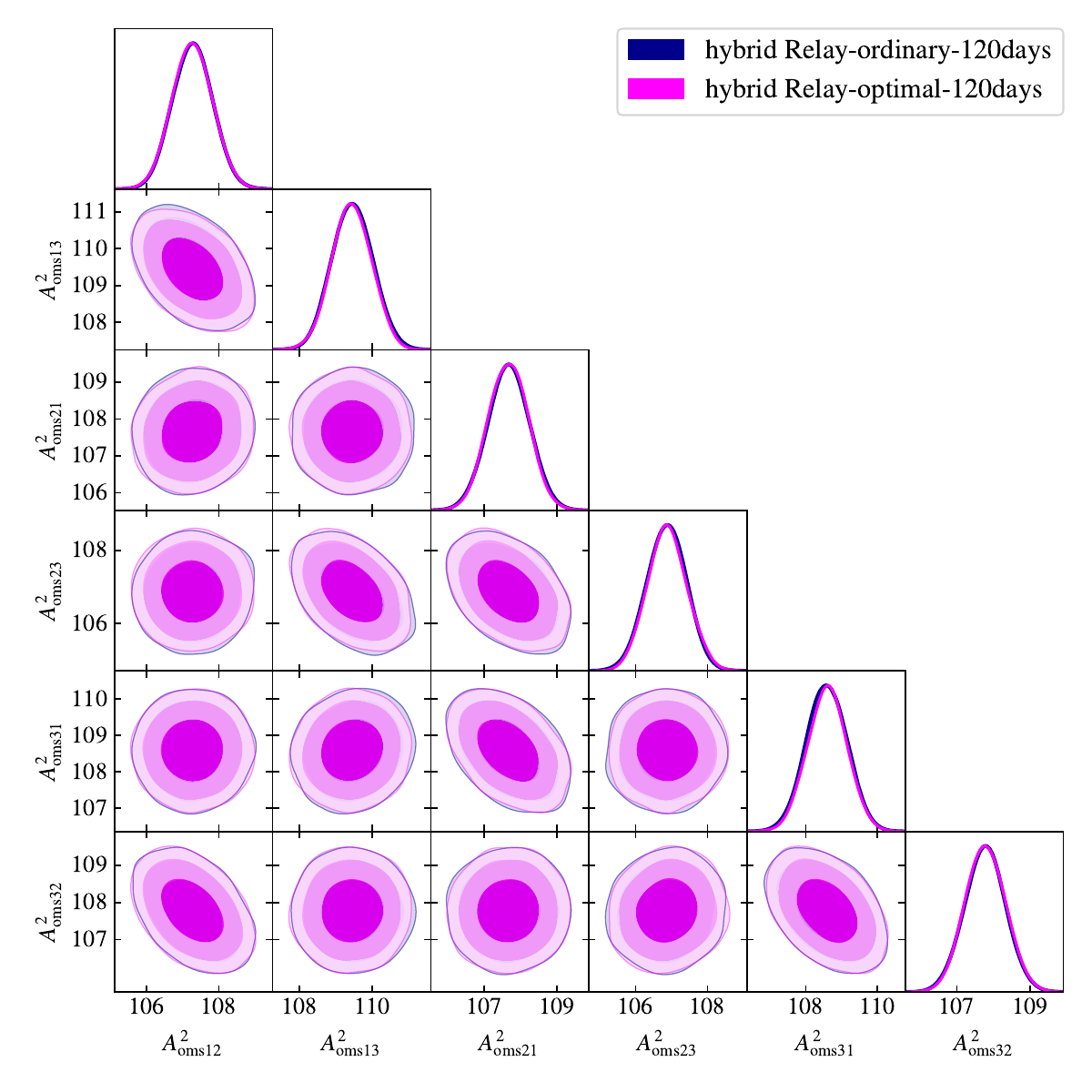}
\caption{\label{fig:corner_hybrid_Relay_ordinary_optimal} The inferred parameter distributions from 120-day ordinary and optimal datasets based on dynamic constellation motion with a 0.03 Hz high frequency cutoff. However, due to the instabilities of T$_{U\overline{U}}$'s spectrum in 120 days duration, the distributions could not infer the noise parameter accurately. }
\end{figure*}

In Section \ref{sec:simulation_characterization}, parameter inferences are performed using optimal observables. As verified in Fig. \ref{fig:corner_hybrid_Relay_ordinary_optimal}, the optimal and ordinary datasets have the identical performances on the noise characterization. 

\section{PSD of null stream $C^{12}_3$} \label{subsec:null_stream_psd}

\begin{figure}[htb]
\includegraphics[width=0.48\textwidth]{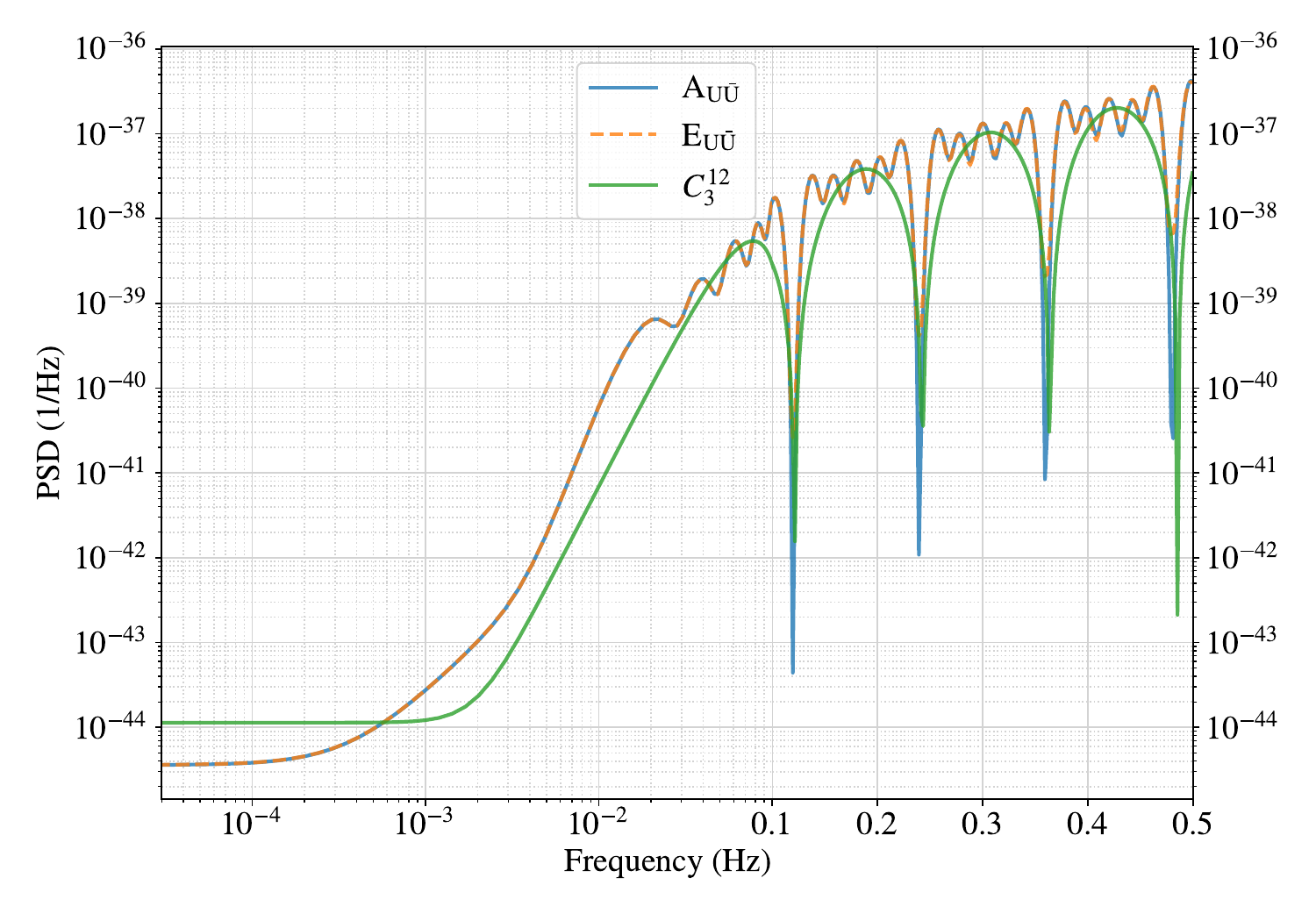}
\caption{\label{fig:psd_AE_zeta12} The noise spectra of A$_\mathrm{U\overline{U}}$, E$_\mathrm{U\overline{U}}$ and $C^{12}_3$ by setting the noise budgets $A_\mathrm{acc}=3$ and $A_\mathrm{oms}=10$ in Eq. \eqref{eq:noise_budgets}. }
\end{figure}

The noise PSD of $C^{12}_3$, including acceleration noise and OMS noise, is expressed as
\begin{equation}
 S_{C^{12}_3} = 96 \sin^4 (\pi f L) S_\mathrm{acc} + 24 \sin^2 (\pi f L) S_\mathrm{oms},
\end{equation}
where $L$ is the arm length. As depicted in Fig. \ref{fig:psd_AE_zeta12}, its noise spectrum only has the same null frequencies as the science channel A$_\mathrm{U\overline{U}}$ and E$_\mathrm{U\overline{U}}$ of hybrid Relay. 

\nocite{*}
\bibliography{apsref}% Produces the bibliography via BibTeX.
%\bibliography{apsempty}% Produces the bibliography via BibTeX.

\end{document}